\documentclass[aps,pre,reprint,
  amsmath, amssymb,
  floatfix,
  superscriptaddress,
  longbibliography,
  nofootinbib]{revtex4-2}

\usepackage{graphicx}
\usepackage{bm}
\usepackage{mathtools}
\usepackage{booktabs}
\usepackage{siunitx}
\usepackage{comment}
   
\newcommand{\bs}[1]{\boldsymbol{#1}}
\newcommand{\CFS}{\Delta\mathrm{CFS}}
\newcommand{\diag}{\mathrm{diag}}
\newcommand{\rkb}{\rho^{(0)}}    
\newcommand{\Nmat}{N}             
\newcommand{\spec}{\rho}          
\DeclareMathOperator{\Cov}{Cov}

\newcommand{\ssch}{\mathrm{SS}}   
\newcommand{\Nch}{\mathrm{N}}     
\newcommand{\Rch}{\mathrm{R}}     

\begin{document}

\title{Non-normal amplification in multitype Hawkes-ETAS models of earthquake triggering}

\author{Didier Sornette}
\email{dsornette@ethz.ch}
\affiliation{Risks-X, Southern University of Science and Technology (SUSTech), Shenzhen, China}

\date{\today}

\begin{abstract}
Earthquake triggering is conventionally characterised by a scalar ETAS branching ratio. We develop a three-type Hawkes-ETAS model that resolves strike-slip, normal, and reverse/thrust earthquakes through a directed branching matrix $\Nmat$. While the spectral radius $\rho(\Nmat)<1$ controls asymptotic stability, its eigenvector geometry controls finite-generation dynamics. Asymmetric cross-mechanism pathways can render $\Nmat$ non-normal, producing large transient and cumulative cascade responses in a strictly subcritical process. We motivate this geometry from receiver-fault availability, Coulomb stress projection, mechanism-dependent magnitude distributions, tectonic loading, and near-degenerate self-triggering. A physically reduced parametrisation separates diagonal self-triggering, a dominant tectonic driver column, and weaker secondary couplings. Numerical examples show that cascade amplification can increase strongly while the eigenvalues remain fixed. Five tectonically informed scenario matrices illustrate plausible geometries. The theory produces six falsifiable predictions for mechanism-resolved catalogues and identifies how scalar ETAS fits may absorb multitype amplification into an apparently elevated branching ratio.
\end{abstract}

\maketitle

\section{Introduction}\label{sec:intro}

Self-exciting (Hawkes) point processes~\cite{Hawkes1971,DaleyVereJones2003,
HawkesOakes1974} have become the dominant statistical framework for
modelling earthquake catalogues since the introduction of the
Epidemic-Type Aftershock Sequence (ETAS) model~\cite{Ogata1988,Ogata1998,
HelmstetterSornette2003a,HelmstetterSornette2002}. In its simplest scalar
form an ETAS model collapses the entire seismic catalogue onto a single
counting process and represents triggering by a magnitude-weighted Omori
kernel complemented by a separable spatial Green function. The scalar branching ratio~$n$, defined as the average number of direct
aftershocks per event integrated over the magnitude distribution, is
the unique parameter governing the stability of the cascade: $n \leq 1$ yields
finite clusters, $n> 1$ explosive supercriticality.

A scalar description is nevertheless too coarse for the physics of earthquake interactions. Earthquakes are commonly classified into three principal focal-mechanism families: strike-slip, normal, and reverse or thrust faulting. These mechanisms correspond to distinct orientations of the stress field and distinct modes of crustal deformation \cite{Scholz2019Mechanics}, but they are not dynamically isolated. Modern moment-tensor catalogues show that multiple focal mechanisms frequently coexist within the same tectonic region and may even be activated within a single earthquake sequence~\cite{Ekstrom2012GCMT,YangHaukssonShearer2012, ChengHauksson2023,RCMT_Italy}. Complex ruptures may also combine several slip modes along different fault segments. This mechanism diversity is therefore not merely a catalogue classification: it is an intrinsic component of the dynamics of crustal deformation. 

The coexistence of several rupture modes is also required by geometric and kinematic constraints. A crustal volume containing faults of different orientations cannot, in general, accommodate an imposed regional deformation by repeated slip of the same focal mechanism. Compatibility of the deformation field requires the activation of complementary fault orientations and slip modes. Geometrical incompatibilities in a fault network accumulate elastic stress that must eventually be released through motion on additional, generally non-coplanar fault systems~\cite{GabrielovKeilisBorokJackson1996}. Thus, even in a region dominated by one tectonic style, subordinate normal, reverse, or strike-slip events are expected to participate in accommodating the total deformation. A mechanism-resolved description is consequently more natural than a scalar model in which all earthquakes belong to a single population. 

The three focal-mechanism families moreover have different triggering
abilities, as indicated by systematic differences in aftershock
productivity, temporal decay, and spatial organisation among
strike-slip, normal, and reverse mainshocks \cite{HoangTrongRouland1993,TahirGrasso2015,DascherCousineauEtAl2020}.
Faulting style influences the spatial and temporal organisation of aftershock sequences, while the associated stress regimes and magnitude distributions affect the productivity and composition of the resulting cascades \cite{TahirGrasso2015,SchorlemmerWiemerWyss2005, PetruccelliEtAl2019b,Hardebeck2014}. Their source radiation patterns, receiver-fault populations, Coulomb stress projections, magnitude distributions, and frictional responses are different. An event of mechanism $j$ may therefore trigger events of mechanism $i$ with an efficiency that differs from that of the reverse pathway. This asymmetry is supported by observations that aftershock productivity depends on the focal mechanism of the parent event~\cite{Hardebeck2014,TahirGrasso2015,BachmannEtAl2011}. Earthquake triggering should thus be represented not by a single branching ratio but by a matrix $\Nmat_{ij}$ describing the expected number of direct offspring of mechanism $i$ produced by an event of mechanism $j$, with $i,j\in\{\ssch,\Nch,\Rch\}$ denoting strike-slip, normal, and reverse events. 

Scalar ETAS models already provide a successful description of the pronounced space--time clustering of seismicity. The multichannel extension is not intended to replace this mechanism, but to refine it by allowing different focal-mechanism populations to have distinct productivities and asymmetric cross-triggering pathways. This added structure can generate a richer and potentially more faithful description of the composition, migration, and intermittency of earthquake clusters.  A rupture on the tectonically dominant mechanism can activate a secondary mechanism on favourably oriented receiver faults; this secondary population can in turn activate a third mechanism or feedback onto the original one. Such directed triggering chains produce episodes in which seismic activity is successively transferred between fault families and spatially distinct fault populations. The resulting bursts need not reflect a globally near-critical or supercritical crust. They may instead arise from finite-generation amplification along particularly efficient sequences of cross-mechanism triggering. After these pathways are exhausted, the activity decays, producing alternating intervals of intense clustered seismicity and relative quiescence. 

The distinction between single and multi-channel triggering is important because observable measures of clustering, including seismicity rates, burst-size distributions, and the apparent scalar-ETAS branching ratio $n_{\mathrm{app}}$, are not controlled solely by the asymptotic stability of the process~\cite{HelmstetterSornette2003a, HelmstetterEtAl2004}. A three-channel model can generate large, intermittent cascades concentrated in particular mechanisms, locations, and time intervals even when the complete process remains asymptotically subcritical. It therefore offers a natural framework for describing both the mechanism composition of earthquake sequences and their spatio-temporally localised bursts. 

Mathematically, this behaviour is associated with a ``non-normal'' branching matrix in the following sense. Because the triggering efficiency from mechanism $j$ to mechanism $i$ need not equal that of the reverse pathway, the eigenvectors of the branching ratio matrix $\Nmat$ are generally non-orthogonal. Perturbations associated with different focal-mechanism populations can then interfere constructively over a finite number of triggering generations, producing \emph{transient amplification} before the cascade ultimately decays (appendix~\ref{sec:nonnormal}). Analogous amplification is well known in hydrodynamic stability and non-normal operator theory~\cite{TrefethenEmbree2005,FarrellIoannou1996,Schmid2007}. Its magnitude is governed not only by the spectral radius $\rho(\Nmat)$, which is the multidimensional generalisation of the scalar branching ratio, but also by the geometry of the eigenvectors, quantified for a diagonalisable matrix by 
the condition number $\kappa(V)=\|V\|\|V^{-1}\|$. 

The aim of this article is therefore to replace the scalar view of seismic branching by a mechanism-resolved theory of statistical seismic flow. The central question is not merely how close to criticality as measured by $\rho(\Nmat)<1$ is the branching matrix \cite{Nandan2021noncrit,LiWuZhuangJiangSornette2025}, but what structure earthquake physics imposes on its entries, eigenvectors, and finite-time amplification properties. We examine whether fault geometry, kinematic compatibility, Coulomb stress transfer, tectonic loading, and rate-and-state friction systematically generate asymmetric and non-normal couplings. In this interpretation, $\kappa(V)$ quantifies the capacity of the physical network of focal-mechanism interactions to generate transient seismic bursts, apparent near- or above-criticality, mechanism switching, and spatio-temporal intermittency while remaining asymptotically subcritical.

We develop and test the multichannel ETAS hypothesis in four stages. First, section~\ref{sec:etaswtrb} introduces the three-channel Hawkes--ETAS model and defines the branching matrix $\Nmat$ governing within- and cross-mechanism triggering. Second, section~\ref{sec:physical} derives structural constraints on $\Nmat$ from five distinct ingredients of earthquake physics: Andersonian constraints on the availability of receiver faults, asymmetry of Coulomb stress projection, mechanism-dependent $b$-values, the preferred direction imposed by tectonic loading and near-degenerate self-triggering. Each ingredient breaks, or reinforces the breaking of, the symmetries that would make $\Nmat$ normal, that is, unitarily diagonalisable with orthogonal eigenvectors. Acting together, they generically favour a non-normal, and in limiting cases near-defective, branching matrix. This provides a physical origin for a large eigenvector condition number $\kappa(V)$ and hence for finite-generation amplification not characterised by the scalar branching ratio alone. 

Third, section~\ref{sec:physical:param} translates these qualitative constraints into a physically informed parametrisation of $\Nmat$. Numerical illustrations are used to examine how the strength and directionality of cross-mechanism coupling control the spectrum, eigenvector geometry, and transient gain, and to formulate falsifiable predictions for mechanism-resolved earthquake sequences. Finally, section~\ref{sec:empirical} constructs empirical proxies for $\Nmat$ from published focal-mechanism statistics in five tectonic settings: the 2011 Tohoku--Oki sequence \cite{AsanoEtAl2011,Hasegawa2011,Hasegawa2012,ObanaEtAl2012}, the 2008 Wenchuan thrust sequence \cite{Wang2009Wenchuan,ZhangEtAl2009Wenchuan, LinEtAl2018Wenchuan}, Southern California strike-slip seismicity \cite{YangHaukssonShearer2012,ChengHauksson2023, YangHauksson2013}, the 2016 Amatrice--Visso--Norcia normal-faulting sequence in central Italy \cite{ChiaraluceEtAl2017,ImprotaEtAl2019,Michele2020Italy}, and the 2019 Ridgecrest orthogonal strike-slip sequence \cite{HaukssonEtAl2019Ridgecrest,WangDregerEtAl2020}. 

These case studies are used to confront four predictions of the theory: driver-column dominance, asymmetric cross-mechanism triggering, amplification of scalar ETAS branching estimates, and enhanced transient amplification when mechanisms other than the dominant driver are activated. They also motivate a fifth, dynamical prediction: major ruptures should transiently increase $\kappa(V)$ by activating and mixing different rupture modes in the post-mainshock stress field, after which $\kappa(V)$ should relax as rate-and-state aftershock transients decay and tectonic loading progressively restores the background mechanism distribution.

The remainder of the paper is structured as follows.
Section~\ref{sec:etaswtrb} introduces the three-channel Hawkes-ETAS model
and reviews multidimensional Hawkes processes.
Their properties are summarised in appendix  \ref{sec:nonnormal}.
Section~\ref{sec:physical} develops the central physical argument of the origin 
of non-normality in the three-channel Hawkes-ETAS model.
Section~\ref{sec:physical:param} presents the physically informed parametrisation of
the branching matrix $N$.
Section~\ref{sec:empirical} presents tentative, physically informed reconstructions from published focal-mechanism statistics, intended as proofs of concept and as roadmaps for future statistically calibrated analyses. Section~\ref{sec:conclusion} concludes. The first appendix
summarises the theoretical framework of non-normal amplification: spectral geometry,
resolvent bounds, generation-operator transient amplification, scalar
reduction bias, and second-order moment amplification. The second and third appendices provide, respectively, a detailed derivation of the resolvent bound and a sensitivity analysis with respect to $\varepsilon_{\mathrm{back}}$, the parameter controlling the weak cross-mechanism feedback from secondary focal mechanisms into the dominant mechanism and the remaining channels.

\section{A three-dimensional Hawkes-ETAS model for earthquake faulting mechanisms}\label{sec:etaswtrb}

\subsection{Multidimensional Hawkes processes: background and stability}\label{sec:hawkes}

We briefly summarise the background needed for what follows; for further
detail and proofs, we refer to Refs.~\cite{Hawkes1971,DaleyVereJones2003,
BremaudMassoulie1996,EmbrechtsLinigerLin2011}.

\subsubsection{Definition and assumptions}\label{sec:hawkes:def}

We consider a $d$-dimensional linear Hawkes process
$\bs{N}(t)=(N_1(t),\dots,N_d(t))^{\top}$ defined on a filtered probability
space $(\Omega,\mathcal{F},(\mathcal{F}_t)_{t\ge 0},\mathbb{P})$, where
each $N_i(t)$ is a simple counting process adapted to
$(\mathcal{F}_t)$. The predictable conditional intensity vector
$\bs{\lambda}(t)=(\lambda_1(t),\dots,\lambda_d(t))^{\top}$ is defined
componentwise by
\begin{equation}
\lambda_i(t)
\;=\;
\mu_i + \sum_{j=1}^{d}\int_{0}^{t}\phi_{ij}(t-s)\,dN_j(s),
\quad i=1,\dots,d,
\label{eq:intensity}
\end{equation}
where $\bs{\mu}=(\mu_1,\dots,\mu_d)^{\top}\in\mathbb{R}_+^d$ is the
background-intensity vector, $\phi_{ij}\colon\mathbb{R}_+\to\mathbb{R}_+$
are the excitation kernels, and the integral is a Stieltjes integral with
respect to the counting measure of $N_j$. Compactly,
\begin{equation}
\bs{\lambda}(t)
\;=\;
\bs{\mu} + \int_0^t \Phi(t-s)\,d\bs{N}(s),
\quad
\Phi(t)=[\phi_{ij}(t)]_{i,j=1}^{d}.
\label{eq:intensity-vec}
\end{equation}
Throughout we assume:
\begin{enumerate}
\item[(i)] (\emph{Non-negativity}) $\phi_{ij}(t)\ge 0$ for all $t\ge 0$
and all $(i,j)$;
\item[(ii)] (\emph{Integrability}) $\int_0^{\infty}\phi_{ij}(t)\,dt<\infty$
for all $(i,j)$;
\item[(iii)] (\emph{Local boundedness}) $\Phi$ is locally integrable on
$\mathbb{R}_+$.
\end{enumerate}

\subsubsection{Integrated branching matrix and Galton-Watson representation}\label{sec:hawkes:branching}

We define the \emph{integrated branching matrix}
\begin{equation}
\Nmat \;=\; \int_0^{\infty}\Phi(t)\,dt,
\qquad
\Nmat_{ij}\;=\;\int_0^{\infty}\phi_{ij}(t)\,dt.
\label{eq:Nmat}
\end{equation}
Each entry $\Nmat_{ij}$ admits the branching interpretation
\begin{equation}
\Nmat_{ij}
\;=\;
\mathbb{E}\bigl[\,\text{\# direct type-}i\text{ offspring of a type-}j\text{ event}\,\bigr].
\label{eq:Nij-interpretation}
\end{equation}
Under (i)-(iii), the Hawkes process admits an equivalent representation as
a multitype Poisson cluster process in which $\Nmat$ plays the role of the
offspring mean matrix of a $d$-type Galton-Watson branching
process~\cite{HawkesOakes1974,DaleyVereJones2003}. Each background event
is the root of an independent cluster; clusters are independent.

\subsubsection{Stationarity and non-explosion}\label{sec:hawkes:stab}

Let $\spec(\Nmat)$ denote the spectral radius of $\Nmat$
(largest eigenvalue modulus). Under (i)-(iii), a stationary version of
the Hawkes process exists and is unique~\cite{BremaudMassoulie1996} if
and only if
\begin{equation}
\spec(\Nmat) \;<\; 1.
\label{eq:stability}
\end{equation}
If $\spec(\Nmat)<1$, the multitype Galton-Watson branching process is
subcritical and clusters have finite expected total progeny; if
$\spec(\Nmat)> 1$, the process is supercritical and no
stationary regime exists. In the scalar case $d=1$, this reduces to the
classical branching-ratio condition $n=\int_0^{\infty}\phi(t)\,dt<1$ of
Refs.~\cite{HawkesOakes1974,HelmstetterSornette2003a}.
The critical case $\spec(\Nmat)=1$ requires separate treatment, since the usual subcritical stationarity conditions 
cease to hold and the process may display qualitatively different scaling behaviour \cite{SaichevSornette2014,Sornette2026CriticalHawkes}.

The stationary mean intensity vector satisfies
\begin{equation}
\bar{\bs{\lambda}} \;=\; (I-\Nmat)^{-1}\bs{\mu}.
\label{eq:stationary-mean}
\end{equation}
This identity makes explicit the central role of the \emph{resolvent}
$(I-\Nmat)^{-1}$ in determining how exogenous loading $\bs{\mu}$ is
amplified by endogenous self- and cross-triggering. The eigenstructure of
$\Nmat$ controls this amplification in ways that go strictly beyond
$\spec(\Nmat)$, as we now show.

\subsection{Formulation of the three-dimensional Hawkes-ETAS model for earthquake faulting mechanisms}\label{sec:etas}

\subsubsection{Channels and marks}

Earthquake catalogues with focal-mechanism information classify each event
by its faulting style: strike-slip ($\ssch$), normal ($\Nch$), or
reverse/thrust ($\Rch$). These three classes are the Andersonian
endmembers~\cite{Anderson1951,Frohlich1992} of a continuous space of
mechanisms parametrised by the rake angle $\lambda$ on the auxiliary
plane. As a first simplifying step, we consider that $|\lambda|\le 30^{\circ}$ defines pure strike-slip,
$\lambda\in[-150^{\circ},-30^{\circ}]$ defines normal, and
$\lambda\in[30^{\circ},150^{\circ}]$ defines reverse, in the Frohlich
ternary partition~\cite{Frohlich1992}. In reality,
many earthquakes have oblique or genuinely mixed focal mechanisms rather than belonging to a single pure endmember; when required, such an event can be represented as two or three coincident elementary events whose strike-slip, normal, and reverse moment tensors are weighted so that their sum reproduces the original seismic moment tensor, as discussed further in Sec.~\ref{sec:empirical:caveats}.
We take
\begin{equation}
i=1\to\ssch,\quad i=2\to\Nch,\quad i=3\to\Rch
\end{equation}
as our three channel indices. The choice of three coarse-grained classes
involves a loss of information that we discuss in
Sec.~\ref{sec:empirical:caveats}, but it is the natural minimal partition
that respects the Andersonian fault-mechanics framework.

Each earthquake is characterised by a mark $(m,\vartheta)$ where $m$ is
the magnitude and $\vartheta$ denotes the geometric parameters of the
focal mechanism (strike, dip, rake) on the chosen nodal plane.

\subsubsection{Conditional intensity}

Let $(t_k,\bs{x}_k,m_k,j_k,\vartheta_k)_{k\ge 1}$ denote the observed
seismic catalogue, where $t_k$ is the occurrence time,
$\bs{x}_k\in\mathbb{R}^2$ the epicentral location, $m_k$ the magnitude,
$j_k\in\{1,2,3\}$ the focal-mechanism class, and $\vartheta_k$ the focal
geometry. The conditional intensity of earthquakes of mechanism $i$ at
$(\bs{x},t)$ is
\begin{equation}
\lambda_i(\bs{x},t)
\;=\;
\mu_i(\bs{x})
+\sum_{k:t_k<t}\phi_{ij_k}(\bs{x}-\bs{x}_k,t-t_k;m_k,\vartheta_k).
\label{eq:cond-int}
\end{equation}
Following standard ETAS practice, we assume the triggering kernel
factorises into magnitude-productivity, temporal, and spatial components:
\begin{equation}
\phi_{ij}(\bs{r},\tau;m,\vartheta)
\;=\;
K_{ij}(m)\,g_{ij}(\tau)\,s_{ij}(\bs{r}\,|\,m,\vartheta),
\label{eq:kernel-fact}
\end{equation}
with $\bs{r}=\bs{x}-\bs{x}_k$, $\tau=t-t_k$, and normalisations
\begin{equation}
\int_0^{\infty}g_{ij}(\tau)\,d\tau\;=\;1,
\quad
\int_{\mathbb{R}^2}s_{ij}(\bs{r}\,|\,m,\vartheta)\,d\bs{r}\;=\;1.
\end{equation}
Under these normalisations, the expected number of direct type-$i$
offspring of a type-$j$ parent of magnitude $m$ is exactly $K_{ij}(m)$.

\subsubsection{Magnitude productivity, Omori decay, and spatial kernel}

Empirical aftershock studies show that productivity scales approximately
exponentially in magnitude~\cite{Felzer2002Productivity,
HelmstetterSornette2003a,Hainzl2008,VanderElstShaw2015}. We therefore
adopt the standard parametrisation
\begin{equation}
K_{ij}(m)\;=\;\kappa_{ij}\,e^{\alpha_{ij}(m-M_0)},
\label{eq:Kij}
\end{equation}
where $M_0$ is the lower magnitude of triggering \cite{SornetteWerner2005}. The coefficient
$\kappa_{ij}$ controls the baseline strength of triggering from
mechanism~$j$ to mechanism~$i$; the exponent $\alpha_{ij}$ governs the
sensitivity to magnitude.

The temporal decay follows a modified Omori law~\cite{Omori1894,Utsu1995,
Ogata1988},
\begin{equation}
g_{ij}(\tau)\;=\;\frac{(p_{ij}-1)\,c_{ij}^{p_{ij}-1}}{(\tau+c_{ij})^{p_{ij}}},
\quad p_{ij}>1,\,c_{ij}>0.
\label{eq:Omori}
\end{equation}

The spatial distribution of triggered events is modelled by a heavy-tailed
isotropic kernel~\cite{Ogata1998,ZhuangEtAl2002}
\begin{equation}
s_{ij}(\bs{r}\,|\,m)
\;=\;
\frac{q_{ij}-1}{\pi\,d_{ij}(m)^2}
\left(1+\frac{\|\bs{r}\|^2}{d_{ij}(m)^2}\right)^{-(1+q_{ij})/2},
~ q_{ij}>0,
\label{eq:spatial}
\end{equation}
with magnitude-scaled characteristic length
$d_{ij}(m)=d_{0,ij}\,e^{\gamma_{ij}(m-M_0)}$.

\subsubsection{Integrated branching matrix}\label{sec:etas:Nmat}

Let $f_j(m)$ denote the magnitude distribution for events of mechanism
$j$. Assuming a Gutenberg-Richter law~\cite{GutenbergRichter1944,
Frohlich1993}
\begin{equation}
f_j(m)\;=\; b_j\ln 10\cdot 10^{-b_j(m-M_0)},\quad m\ge M_0,
\end{equation}
the integrated branching matrix entry is
\begin{equation}
\Nmat_{ij}
\;=\;
\mathbb{E}_{m\sim f_j}\bigl[K_{ij}(m)\bigr]
\;=\;
\kappa_{ij}\,\mathbb{E}_{m\sim f_j}\bigl[e^{\alpha_{ij}(m-M_0)}\bigr].
\end{equation}
The expectation exists provided $\alpha_{ij}<\beta_j:= b_j \ln(10)$, in which
case
\begin{equation}
\Nmat_{ij}
\;=\;
\frac{\kappa_{ij}}{1-\alpha_{ij}/\beta_j}.
\label{eq:Nij-formula}
\end{equation}
This is the central object of the present paper. The diagonal entries
$\Nmat_{ii}$ describe self-triggering within a mechanism; the off-diagonal
entries $\Nmat_{ij},\,i\ne j$, describe cross-mechanism triggering. Since
$\Nmat$ is in general non-normal (see appendix \ref{sec:nonnormal}), its eigenvectors are not orthogonal,
and collective triggering modes may exhibit transient amplification even
when $\spec(\Nmat)<1$.

We are now in a position to ask the central question: what does the
physics of earthquake rupture, stress transfer, fault mechanics, and
tectonic loading dictate for the form of $\Nmat$?

\section{Physical origin of the non-normality of the branching matrix $N$}  \label{sec:physical}

The non-normal amplification analysed in Appendix~\ref{sec:nonnormal} is controlled by the eigenvector condition number $\kappa(V)$ of the branching matrix $\Nmat$. When $\Nmat$ is normal, its eigenvectors may be chosen orthonormal and $\kappa(V)=1$. Values $\kappa(V)>1$ therefore quantify the departure from normality, with large values indicating strongly non-orthogonal eigenvectors and proximity, in limiting cases, to a defective matrix. It is thus essential to ask: \emph{which mechanical processes can generate the asymmetric structure of $\Nmat$ responsible for large $\kappa(V)$ and the associated finite-time amplification?}
We argue that the
physics of static and dynamic stress transfer, Andersonian fault
mechanics, magnitude-distribution asymmetry, and tectonic loading
hierarchies do not permit $\Nmat$ to be normal. They contribute five
mechanisms that, taken together, force $\Nmat$ into a near-rank-one,
near-defective form. We \emph{do not} postulate the form of $\Nmat$ a
priori; rather we derive successive constraints from first principles and
show that the resulting object is generically far from normal, with
$\kappa(V)$ controlled by a small number of identifiable physical
parameters.

\subsection{Mechanistic factorisation}\label{sec:physical:factor}

For each ordered pair $(i,j)$ of receiver and source mechanisms, the branching coefficient $\Nmat_{ij}$ is the expected number of direct type-$i$ offspring generated by a type-$j$ parent event. We decompose $\Nmat_{ij}$ into three factors associated with distinct physical ingredients: 
\begin{equation} 
\Nmat_{ij} = \underbrace{\rkb_i}_{\text{receiver availability}} \cdot \underbrace{G_{ij}}_{\text{stress-transfer gain}} 
\cdot \underbrace{\eta_j}_{\text{source productivity}} . 
\label{eq:factorisation} 
\end{equation} 
The three factors are:
\begin{itemize}
\item $\rkb_i\ge 0$: the probability density that a randomly oriented
fault patch in the volume satisfies the Andersonian/Mohr-Coulomb
criterion for mechanism class $i$ under the background stress tensor
$\bs{\sigma}^{(0)}$. This depends only on the background stress and on
the regional distribution of preexisting fault orientations.

\item $G_{ij}\ge 0$: the static and dynamic stress-transfer plus
rate-and-state frictional gain, which is the expected coseismic increase in
seismicity rate at a type-$i$ receiver per unit moment of a type-$j$
source, integrated over the spatial kernel
$s_{ij}(\bs{r}\,|\,m,\vartheta)$.

\item $\eta_j$: the magnitude-integrated source productivity of mechanism
$j$, appearing in the denominator of~(\ref{eq:Nij-formula}):
$\eta_j\propto[1-\alpha_j/\beta_j]^{-1}$.
\end{itemize}
Each factor imposes a different structural constraint on $\Nmat$. We
examine them in turn.

\subsection{Andersonian receiver availability forces row-rank reduction}\label{sec:physical:Anderson}

In a region of fixed tectonic regime, the background stress tensor
$\bs{\sigma}^{(0)}$ has a well-defined orientation. Anderson's
classification~\cite{Anderson1951} identifies three endmembers
characterised by which of the principal stress axes is vertical:
\begin{equation}
\begin{aligned}
&\text{Normal regime:}    && \sigma_1 \text{ vertical},\\
&\text{Strike-slip regime:}&& \sigma_2 \text{ vertical},\\
&\text{Reverse regime:}   && \sigma_3 \text{ vertical},
\end{aligned}
\end{equation}
where $\sigma_1>\sigma_2>\sigma_3$ are the principal compressive stresses.

Concretely, for stress ratio $R=(\sigma_1-\sigma_2)/(\sigma_1-\sigma_3)$
and optimal Mohr-Coulomb angle $\theta_c=\frac{\pi}{4}+\frac{1}{2}\arctan(\mu_f)$
with $\mu_f$ the static friction coefficient, the favoured fault
orientations form two narrow conjugate planes whose Andersonian class is
unique. The dispersion away from the favoured class is controlled by
stress heterogeneity and by inherited fault populations.
Empirically, the Wallace-Bott
hypothesis~\cite{Wallace1951,Bott1959,Celerier2008} that slip is parallel
to the projected shear traction on the receiver provides a quantitative
constraint: the admissible rake on a receiver is fully determined by
$\bs{\sigma}^{(0)}$ once the orientation is fixed, so heterogeneity in
$\rkb$ is bounded by the heterogeneity in $\bs{\sigma}^{(0)}$.

Hardebeck's analysis of the southern California
catalogue~\cite{Hardebeck2014} found that approximately $10$-$15\%$ of
aftershocks are ``poorly oriented outliers'' relative to the
background-stress prediction. Operationally, this means that, in a
fixed-regime sub-region of the crust, the vector
$\rkb=(\rkb_{\ssch},\rkb_{\Nch},\rkb_{\Rch})^{\top}$ is dominated by one
component, say $\rkb_{i_\star}\gg \rkb_{i\ne i_\star}$, with the
subdominant components representing the $\sim 10\%$ outlier fraction.

The entry-wise factorisation~(\ref{eq:factorisation}) is just the matrix
product
\begin{equation}
\Nmat \;=\; \diag(\rkb)\,G\,\diag(\bs{\eta}),
\label{eq:Nmat-decomp}
\end{equation}
in which $\diag(\rkb)$ pre-multiplies $G$ row-wise and $\diag(\bs{\eta})$
post-multiplies it column-wise. The Andersonian argument above states that
$\rkb$ is \emph{row-sparse}: in a single-regime crustal volume only the
component $\rkb_{i_\star}$ associated with the regime-favoured receiver
class is appreciable, while $\rkb_{i\ne i_\star}=O(\varepsilon_{\text{out}})$
with $\varepsilon_{\text{out}}\sim 0.10$-$0.15$ estimated from the
Hardebeck outlier fraction.

Row-sparsity of $\rkb$ has an immediate consequence for the rank
structure of $\Nmat$. Pre-multiplication of $G$ by the rank-1-plus-small
diagonal matrix $\diag(\rkb)$ produces a matrix one of whose rows
(the $i_\star$-th) is of order $O(1)$ while the other two rows are of order
$O(\varepsilon_{\text{out}})$. In the strict limit
$\varepsilon_{\text{out}}\to 0$, only the $i_\star$-th row of $\Nmat$ is
non-zero and $\Nmat$ is exactly rank one; for small but finite
$\varepsilon_{\text{out}}$ the matrix is rank three but its singular
values are split as
$\sigma_1\!=\!O(1)\gg\sigma_2,\sigma_3=O(\varepsilon_{\text{out}})$, i.e.\
its \emph{effective} rank in the sense of the numerical-rank threshold is
one. \emph{Andersonian receiver mechanics therefore generates row-rank
reduction, the dominant source of small effective rank in $\Nmat$ in
regime-coherent crustal regions.}

\subsection{Coulomb projection asymmetry forces $G_{ij}\ne G_{ji}$}\label{sec:physical:Coulomb}

A coseismic slip event of mechanism $j$ produces a static stress
perturbation $\Delta\bs{\sigma}^{(j)}(\bs{r})$, a symmetric rank-two
tensor~\cite{KingSteinLin1994,Stein1999}. The Coulomb failure stress
change~\cite{Reasenberg1992,KingSteinLin1994,ToddStein2003} projected onto
a receiver of mechanism $i$ with normal $\hat{\bs{n}}_i$ and slip
direction $\hat{\bs{s}}_i$ is
\begin{equation}
\CFS_{i\leftarrow j}(\bs{r})
\;=\;
\hat{\bs{s}}_i^{\top}\Delta\bs{\sigma}^{(j)}(\bs{r})\,\hat{\bs{n}}_i
\;+\;
\mu'\,\hat{\bs{n}}_i^{\top}\Delta\bs{\sigma}^{(j)}(\bs{r})\,\hat{\bs{n}}_i,
\label{eq:CFS}
\end{equation}
where $\mu'=\mu_f(1-B)$ is the effective friction with Skempton's
coefficient $B$. The projection map
$\mathbb{P}_i\colon\bs{\sigma}\mapsto\hat{\bs{s}}_i^{\top}\bs{\sigma}\,\hat{\bs{n}}_i+\mu'\hat{\bs{n}}_i^{\top}\bs{\sigma}\,\hat{\bs{n}}_i$
is linear but \emph{not symmetric in the index pair $(i,j)$}: even when
$\Delta\bs{\sigma}^{(j)}=\Delta\bs{\sigma}^{(i)}$ (which never holds in
practice), swapping $(i,j)$ replaces $\mathbb{P}_i$ by $\mathbb{P}_j$,
which generally returns a different number because $\hat{\bs{n}}_i\ne\hat{\bs{n}}_j$ and $\hat{\bs{s}}_i\ne\hat{\bs{s}}_j$.

Concrete asymmetries imposed by the geometry of static stress transfer
include:
\begin{itemize}
\item \emph{Subduction megathrust $\to$ outer-rise normal.} A
low-angle thrust source on the megathrust unclamps the down-going slab in
the outer-rise region: $\Delta\sigma_n<0$ for normal-fault receivers in
that volume, yielding large positive $\CFS_{\Nch\leftarrow\Rch}$. The
2011 Tohoku-Oki $M_w\,9.0$ event triggered the $M_w\,7.6$ outer-rise
normal aftershock \SI{39}{\minute} later, plus extensive normal-fault
microseismicity in the subducting Pacific
plate~\cite{ObanaEtAl2012,AsanoEtAl2011,Hasegawa2011}. Conversely a
normal-faulting event in the slab produces only a marginal load on the
megathrust receiver. The asymmetry
$\Nmat_{\Nch\Rch}\gg\Nmat_{\Rch\Nch}$ is therefore expected to be at
least an order of magnitude at given moment.

\item \emph{Strike-slip rupture $\to$ conjugate strike-slip vs.\
dipping receivers.} A vertical strike-slip slip plane primarily loads
conjugate strike-slip receivers and shadows perpendicular ones; thrust or
normal receivers in the same volume are loaded much more weakly, with
geometric scaling $\sim\cos 2\theta$ vs.\ $\sim\sin 2\theta$ on the Mohr
circle.

\item \emph{Sign reversal and stress shadows.} $G_{ij}>0$ for some
$(i,j)$ pairs and $G_{ij}<0$ for others (the latter correspond to
stress-shadow geometry~\cite{HardebeckHauksson2001Shadow,
Toda2005Animation}). In a productivity-based ETAS formulation $\kappa_{ij}\ge 0$, so negative entries are replaced by zero, introducing a hard
zero pattern that is intrinsically non-symmetric.
\end{itemize}
The degree of asymmetry of $G$ is controlled by the angular mismatch between the principal axes of the stress perturbation $\Delta\bs{\sigma}^{(j)}$ generated by the source and the slip direction of the receiver fault. For receivers belonging to different Andersonian classes, this mismatch is typically of order $30^{\circ}$--$90^{\circ}$. Consequently, the transfer efficiency from mechanism $j$ to mechanism $i$ generally differs from that of the reverse pathway, so that $G$ is generically non-symmetric and may be strongly non-normal; see Appendix~\ref{sec:nonnormal}.

The conversion of Coulomb stress changes into triggering rates may be further modulated by receiver-dependent rate-and-state parameters \cite{Dieterich1994,CoccoRice2002}. This nonlinearity can alter the magnitude of an asymmetry already present in the source--receiver geometry, but it is not by itself a distinct origin of non-normal coupling.

\subsection{The $b$-value asymmetry produces column weighting}\label{sec:physical:bvalue}

Global moment-tensor catalogues indicate that the Gutenberg--Richter $b$-value depends systematically on focal mechanism \cite{SchorlemmerWiemerWyss2005}. Representative estimates satisfy \begin{equation} b_{\Nch} \approx 1.10 \;>\; b_{\ssch} \approx 1.00 \;>\; b_{\Rch} \approx 0.95, \label{eq:b-mechanism} \end{equation} although the precise numerical values depend on the catalogue, magnitude range, depth distribution, and classification procedure. This ordering is commonly interpreted in terms of the inverse dependence of $b$ on differential stress: compressional regimes dominated by reverse or thrust faulting sustain, on average, larger differential stresses than extensional regimes dominated by normal faulting, with strike-slip regimes occupying an intermediate position \cite{SchorlemmerWiemerWyss2005}. 
The dependence on faulting style has subsequently been recovered in regional and mechanism-resolved analyses \cite{GuliaWiemer2010,PetruccelliEtAl2019a,PetruccelliEtAl2019b}. 
More broadly, the earthquake-size distribution also varies among tectonic environments. In particular, a generalized-Pareto analysis of the global Harvard moment-tensor catalogue found a markedly larger $b$-value for mid-ocean-ridge seismicity, $b \simeq 1.50\pm0.09$, than for the other major tectonic zones, $b \simeq 1.00\pm0.05$, with high statistical significance \cite{PisarenkoSornette2003}. In terms of the complementary cumulative seismic-moment distribution, these values correspond approximately to power-law exponents $1$ and $2/3$, respectively. Taken together, these results suggest that the magnitude distribution is not universal, but depends on both focal mechanism and the mechanical and tectonic environment.

Consequently $\beta_j$ in
(\ref{eq:Nij-formula}) is \emph{column-dependent}, and the amplification
denominator $[1-\alpha_{ij}/\beta_j]^{-1}$ is largest for the
column $j$ with smallest $\beta_j := b_j \ln(10)$. Taking for instance $\alpha\simeq 1$, the
relative column weights are
\begin{equation}
\eta_{\Rch}:\eta_{\ssch}:\eta_{\Nch}\;\simeq\;1.84\,:\,1.77\,:\,1.65,
\end{equation}
i.e.\ $\sim 10$-$12\%$ amplification of reverse over normal columns. This
is a column-multiplicative asymmetry, equivalent to right-multiplying
$\Nmat$ by $\diag(\bs{\eta})$. Combined with the row-rank reduction of
Sec.~\ref{sec:physical:Anderson}, $\Nmat$ inherits an explicit non-symmetric
outer-product backbone $\rkb\bs{\eta}^{\top}$ that is rank-one in the
absence of $G$. This is the \emph{structural seed of non-normality}
(see appendix \ref{sec:nonnormal}).

\subsection{Tectonic loading direction makes $G$ near-triangular}\label{sec:physical:tectonic}

Beyond the static stress transfer of a single event, the secular tectonic
loading direction imposes a hierarchical relationship among mechanism
classes. In a convergent regime, the persistent tectonic loading projects predominantly onto reverse or thrust mechanisms, which are progressively reloaded after each rupture. The co-seismic stress field generated by a reverse-faulting event, however, can also activate normal-faulting receivers in the outer rise or back-arc and strike-slip receivers in oblique transfer zones. Tectonic loading and earthquake-mediated stress transfer therefore act through different directional pathways, naturally producing asymmetric cross-mechanism couplings. This produces a directed cascade
\begin{equation}
\Rch \;\longrightarrow\; \{\Nch,\,\ssch\}\quad (\text{convergent}),
\end{equation}
with reverse loading ``upstream'' and the other mechanisms ``downstream''.
Analogous cross-mechanism pathways may occur in extensional and transcurrent regimes. Their strength and even their sign depend on the orientation and location of the receiver faults relative to the source stress perturbation. For example, normal-faulting earthquakes may promote strike-slip or reverse slip on suitably oriented neighbouring structures, while strike-slip earthquakes may promote either normal or reverse faulting in extensional and compressional quadrants of their co-seismic stress field. These possibilities must therefore be evaluated from the source--receiver geometry and Coulomb stress transfer, rather than assumed from the tectonic regime alone.

The cascade should not, however, be read as strictly unidirectional. A
crucial geometric mechanism that loads off-driver mechanisms in any
regime, and that is particularly potent in transcurrent settings, is the
\emph{geometric incompatibility} of non-coplanar fault
systems~\cite{GabrielovKeilisBorokJackson1996}. A rigid-block fault network
embedded in an elastic medium cannot in general accommodate arbitrary
slip on its constituent faults while preserving continuity of the
displacement field: the slip-rate budget around any closed loop of fault
segments, in particular around any triple junction, fails to close in
general, producing an incompatibility tensor that accumulates as stress
on a timescale set by the loading rate and the deviation of the local
fault geometry from planarity. The accumulated incompatibility stress
must be released by off-system seismicity, which generically projects
onto mechanism classes orthogonal to the dominant slip direction of the
network. Textbook examples are thrust pop-ups at restraining bends of
strike-slip faults (e.g.\ the San Bernardino Mountains uplift along the
Big Bend of the San Andreas; the Lebanon ranges along the Dead Sea
transform) and normal pull-aparts at releasing bends (e.g.\ the Salton
Trough; the Dead Sea pull-apart basin
itself~\cite{GarfunkelEtAl1981}. In our framework, these
configurations correspond to non-zero $\Nmat_{\Rch,\ssch}$ and
$\Nmat_{\Nch,\ssch}$ entries in transcurrent regimes, partially
populating the column of $\Nmat$ corresponding to the strike-slip driver.
Analogous incompatibility effects arise at the bends and tips of normal
and reverse fault systems, although, for those mechanisms, the geometric
constraint is less acute because the dominant slip is dip-parallel and
the system can absorb local incompatibility through aseismic creep or
folding more readily than a transcurrent system can.

The net consequence is that $G$ in (\ref{eq:Nmat-decomp}) is
\emph{nearly triangular but not strictly so}: the driver column is the
dominant one, but the geometric-incompatibility mechanism prohibits a
strict-triangular limit and sets a lower bound on the back-coupling
parameter $\varepsilon_{\text{back}}$ in the convergent/extensional
regimes (small but non-zero, $\sim 0.02$-$0.05$) and a larger lower
bound in the transcurrent regime (typically
$\varepsilon_{\text{back}}\sim 0.05$-$0.15$, reflecting the abundance
of bends, step-overs, and triple junctions in continental strike-slip
systems).
When the channels are ordered with the driver mechanism in the last
coordinate, the off-diagonal of $G$ (and hence of $\Nmat$) is heavily
concentrated in the column corresponding to the driver. Up to a weak back-coupling contribution $\varepsilon_{\mathrm{back}}\sim0.02$--$0.10$, associated with elastic feedbacks and minority receiver-fault populations, the branching matrix can be written as \begin{equation} \Nmat = D + \varepsilon_{\mathrm{drive}}\,T_\sigma + \varepsilon_{\mathrm{back}}\,B, \label{eq:N-tri} \end{equation} where $D=\diag(\Nmat_{\ssch\ssch},\Nmat_{\Nch\Nch},\Nmat_{\Rch\Rch})$ contains the within-mechanism branching ratios, $T_\sigma$ represents the dominant directed cross-mechanism pathways imposed by the tectonic loading geometry, and $B$ collects the weaker reverse and secondary couplings. The matrix $T_\sigma$ is strictly triangular and therefore nilpotent: all its eigenvalues vanish, although the matrix itself is generally nonzero. It is consequently non-normal unless it vanishes identically. When added to the nearly degenerate diagonal matrix $D$, this directed triangular coupling strongly rotates and aligns the eigenvectors of $\Nmat$, potentially producing a large condition number $\kappa(V)$. In the limiting case of vanishing back-coupling and exactly degenerate diagonal entries, $\Nmat$ approaches a defective matrix. If the two successive links of the three-channel triggering chain are both nonzero, the limiting nilpotent part forms a Jordan block of size~3, namely a canonical matrix block associated with one eigenvalue but possessing fewer independent eigenvectors than its dimension; otherwise, the largest Jordan block has size~2.

\subsection{Near-degenerate self-triggering enhances non-normality} \label{sec:physical:degen} 

The diagonal entries $\Nmat_{ii}$ quantify the mean number of direct same-mechanism offspring produced by an event of mechanism $i$. Global subcriticality requires $\rho(\Nmat)<1$, but this condition alone places no strong constraint on the relative values of the three diagonal entries. In particular, a mechanism with weak self-triggering may remain well represented in the catalogue through background loading and cross-mechanism triggering. The relevant observation for the present argument is instead conditional: if the within-mechanism branching ratios are comparable, then the directed off-diagonal couplings in Eq.~(\ref{eq:N-tri}) can produce strongly non-orthogonal eigenvectors. To make this explicit, write 
\begin{equation} \Nmat_{ii}=\bar n+\delta_i, \qquad \max_i|\delta_i|\ll \varepsilon_{\mathrm{drive}}\|T_\sigma\|, 
\label{eq:diag-spread} 
\end{equation} 
where $\bar n<1$ is a representative self-triggering level and the $\delta_i$ measure the spread among mechanisms. When the diagonal entries are well separated, the eigenvectors remain comparatively well conditioned despite the asymmetry of the off-diagonal terms. By contrast, when the diagonal entries are close, the directed coupling mixes nearly resonant mechanism channels, and $\Nmat$ can approach a defective limit as the back-coupling tends to zero. The eigenvector condition number $\kappa(V)$ may then become large. Near-degeneracy of the self-triggering terms constitutes an additional condition that can substantially enhance non-normal amplification. Whether this condition is realised in earthquake catalogues must ultimately be tested using mechanism-resolved statistical calibration of the multichannel ETAS model.

\subsection{Synthesis: what physics dictates for $\Nmat$}\label{sec:physical:synthesis}

The five mechanisms~\ref{sec:physical:factor}-\ref{sec:physical:degen}
combine into a falsifiable structural prediction:

\smallskip
\begin{quote}
\emph{$\Nmat$ is the sum of a \textbf{near-degenerate self-triggering diagonal} ($D$, with all $\Nmat_{ii}$ within $\lesssim 0.2$ of one another and each $<1$), an \textbf{upstream-driver triangular block} ($T_\sigma$, with the driver mechanism populating the dominant column and the regional receiver availability selecting the dominant row), and a \textbf{small back-coupling perturbation} ($B$, of order $5$-$15\%$, encoding stress rotation and minority fault populations). The eigenvector condition number $\kappa(V)$ is large precisely when (\ref{eq:diag-spread}) is tight and the driver column is concentrated.}
\end{quote}
\smallskip
Three immediate consequences derive:
\begin{itemize}
\item[(a)] \emph{Non-normality is unavoidable.} Symmetric $\Nmat$ would
require either isotropic receivers (no Andersonian regime), and equal
$b$-value across mechanisms (contradicts~\cite{SchorlemmerWiemerWyss2005}),
and absence of tectonic loading direction (contradicts plate
tectonics~\cite{DemetsEtAl1990}), and breakdown of Coulomb stress geometry
(contradicts a century of fault mechanics). Each constraint independently
violates symmetry; their combination forces $\Nmat\ne\Nmat^{\top}$ with a
definite direction.

\item[(b)] \emph{Near-defective limit is generic.} In a region of
well-defined tectonic regime, all five mechanisms push $\Nmat$ toward the
defective limit: row-rank reduction (Andersonian receiver) gives low
effective rank; triangular off-diagonal (loading direction) gives
Jordan-block geometry; degenerate diagonal (subcriticality) gives
eigenvalue collision.

\item[(c)] \emph{Non-normality is a diagnostic of coupling geometry.} The eigenvector condition number $\kappa(V)$ reflects the organisation and directionality of the interactions among focal-mechanism populations. A large $\kappa(V)$ is expected when triggering is concentrated along a small number of strongly asymmetric pathways, especially when these pathways couple mechanism channels with similar self-triggering rates. More distributed or reciprocal coupling may reduce this alignment, but mixed tectonic settings do not necessarily imply weaker non-normality: depending on their source--receiver geometry, they may either disrupt a dominant triggering hierarchy or generate additional asymmetric cross-mechanism pathways. Thus, $\kappa(V)$ should be interpreted as an empirical diagnostic of the effective coupling structure rather than as a universal classifier of single- versus mixed-regime tectonics; see Appendix~\ref{sec:nonnormal}.
\end{itemize}

\section{Physically informed parametrisation of the branching ratio matrix $N$}\label{sec:physical:param}

\subsection{Formulation and definitions}
\label{sec:physical:param:formulation}

The branching matrix $N$ has already been defined statistically in
Sec.~\ref{sec:etas:Nmat}: its entry $N_{ij}$ is the expected number of
direct type-$i$ offspring generated by a type-$j$ parent, and its
connection with the parameters of the marked Hawkes--ETAS model is given
by Eq.~\eqref{eq:Nij-formula}. Here, we reorganise the same matrix so that its entries
can be interpreted in terms of the physical mechanisms developed in
Sec.~\ref{sec:physical}.

In particular, Eq.~\eqref{eq:factorisation} expresses each already-defined
branching coefficient as
\begin{equation}
N_{ij}=r_iG_{ij}\eta_j,
\label{eq:physical-factorisation-recalled}
\end{equation}
where $r_i$ represents the availability of receivers of mechanism $i$,
$G_{ij}$ the source--receiver stress-transfer gain, and $\eta_j$ the
magnitude-integrated productivity of a parent of mechanism $j$. Equation
\eqref{eq:physical-factorisation-recalled} is therefore a physical
factorisation of the ETAS branching coefficient defined earlier.

Let $j_\star$ denote the parent mechanism associated with the dominant
directed triggering pathway in the tectonic regime under consideration.
For example, $j_\star=\mathrm{R}$ in a reverse/thrust-dominated setting
and $j_\star=\mathrm{SS}$ in a strike-slip-dominated setting. Once
$j_\star$ is chosen, the entries of $N$ can be partitioned exactly into
three non-overlapping parts:
\begin{equation}
N
=
D
+
a_{\mathrm{drv}}\,
\bm{\pi}_{\mathrm{drv}}\bm e_{j_\star}^{\mathsf T}
+
B_{\mathrm{weak}}.
\label{eq:N-reduced}
\end{equation}
The three terms contain, respectively, the diagonal self-triggering
coefficients, the off-diagonal part of the driver-parent column, and all
remaining off-diagonal coefficients. 

The diagonal contribution is
\begin{equation}
D=\operatorname{diag}(n_{\mathrm{SS}},n_{\mathrm{N}},n_{\mathrm{R}}),
\qquad
n_i:=N_{ii}=r_iG_{ii}\eta_i.
\label{eq:N-reduced-diagonal}
\end{equation}
The quantities $n_i$ are the actual within-mechanism branching ratios.
The near-degenerate regime discussed in
Sec.~\ref{sec:physical:degen} corresponds directly to comparable values
of $n_{\mathrm{SS}}$, $n_{\mathrm{N}}$, and $n_{\mathrm{R}}$.

The off-diagonal entries of the driver column are collected in the
rank-one matrix
$a_{\mathrm{drv}}\bm{\pi}_{\mathrm{drv}}
\bm e_{j_\star}^{\mathsf T}$. Its total branching strength is defined by
\begin{equation}
a_{\mathrm{drv}}
:=
\sum_{i\ne j_\star}N_{i j_\star}
=
\eta_{j_\star}
\sum_{i\ne j_\star}r_iG_{i j_\star}.
\label{eq:driver-strength}
\end{equation}
When $a_{\mathrm{drv}}>0$, the corresponding child-mechanism composition
is
\begin{equation}
\pi_{\mathrm{drv},i}
:=
\frac{N_{i j_\star}}{a_{\mathrm{drv}}}
=
\frac{r_iG_{i j_\star}}
{\displaystyle\sum_{k\ne j_\star}r_kG_{k j_\star}},
\qquad i\ne j_\star,
\label{eq:driver-composition}
\end{equation}
with
\begin{equation}
\pi_{\mathrm{drv},j_\star}=0,
\qquad
\sum_{i\ne j_\star}\pi_{\mathrm{drv},i}=1.
\label{eq:driver-composition-normalisation}
\end{equation}
Consequently,
\begin{equation}
a_{\mathrm{drv}}\pi_{\mathrm{drv},i}
=N_{i j_\star}
=r_iG_{i j_\star}\eta_{j_\star},
\qquad i\ne j_\star.
\label{eq:physical-map-driver}
\end{equation}
If $a_{\mathrm{drv}}=0$, the driver term vanishes and
$\bm{\pi}_{\mathrm{drv}}$ need not be defined.

The residual matrix contains the remaining off-diagonal entries:
\begin{equation}
(B_{\mathrm{weak}})_{ij}
:=
\begin{cases}
N_{ij}=r_iG_{ij}\eta_j,
& i\ne j,\quad j\ne j_\star,\\
0,
& i=j\ \text{or}\ j=j_\star.
\end{cases}
\label{eq:physical-map-weak}
\end{equation}
The adjective ``weak'' records the physical hypothesis, developed in
Sec.~\ref{sec:physical:tectonic}, that these reverse and secondary
pathways are smaller than the dominant driver pathway. Their inclusion in
Eq.~\eqref{eq:N-reduced}, however, is exact and does not require them to
be neglected.

This construction makes the relation with the schematic form introduced
in Eq.~\eqref{eq:N-tri} explicit. The diagonal matrix $D$ is unchanged;
the directed term $\varepsilon_{\mathrm{drive}}T_\sigma$ is identified
with the off-diagonal driver column
$a_{\mathrm{drv}}\bm{\pi}_{\mathrm{drv}}
\bm e_{j_\star}^{\mathsf T}$; and the back-coupling term
$\varepsilon_{\mathrm{back}}B$ is represented by
$B_{\mathrm{weak}}$. 

For the ordering
$1=\mathrm{SS}$, $2=\mathrm{N}$, $3=\mathrm{R}$ and a
reverse/thrust driver $j_\star=\mathrm{R}$, the driver-column term is
\begin{equation}
a_{\mathrm{drv}}\bm{\pi}_{\mathrm{drv}}\bm e_3^{\mathsf T}
=
\begin{pmatrix}
0 & 0 & a_{\mathrm{drv}}\pi_{\mathrm{SS}\mid\mathrm{R}}\\
0 & 0 & a_{\mathrm{drv}}\pi_{\mathrm{N}\mid\mathrm{R}}\\
0 & 0 & 0
\end{pmatrix}.
\label{eq:reverse-driver-column}
\end{equation}
The zero in the $(\mathrm{R},\mathrm{R})$ entry is essential: reverse-to-
reverse self-triggering is already contained in $D$ and must not be
counted again in the driver term.

The complete matrix then takes the explicit form
\begin{equation}
N_{\mathrm{R}}
=
\begin{pmatrix}
n_{\mathrm{SS}}
& b_{\mathrm{SS}\leftarrow\mathrm{N}}
& a_{\mathrm{drv}}\pi_{\mathrm{SS}\mid\mathrm{R}}
\\
b_{\mathrm{N}\leftarrow\mathrm{SS}}
& n_{\mathrm{N}}
& a_{\mathrm{drv}}\pi_{\mathrm{N}\mid\mathrm{R}}
\\
b_{\mathrm{R}\leftarrow\mathrm{SS}}
& b_{\mathrm{R}\leftarrow\mathrm{N}}
& n_{\mathrm{R}}
\end{pmatrix},
\label{eq:Nmat-explicit-reverse-driver}
\end{equation}
where
\begin{equation}
b_{i\leftarrow j}:=(B_{\mathrm{weak}})_{ij}
=N_{ij}=r_iG_{ij}\eta_j,
\qquad i\ne j,\quad j\ne \mathrm{R}.
\label{eq:weak-entry-notation}
\end{equation}
The arrow in $b_{i\leftarrow j}$ follows the convention that columns
represent parent mechanisms and rows represent child mechanisms. The
analogous matrices for normal- and strike-slip-driven regimes follow by
permuting the mechanism labels.

Equation~\eqref{eq:N-reduced} does not by itself reduce the number of
independent entries of $N$: it is an organisational decomposition of an
arbitrary nonnegative branching matrix after one driver column has been
selected. A genuine reduction arises only if additional physically
testable restrictions are imposed, for example approximate degeneracy of
the $n_i$, dominance of the driver-column coefficients over the entries
of $B_{\mathrm{weak}}$, or a lower-dimensional model for the residual
couplings. The advantage of the formulation is therefore conceptual and
empirical: it separates self-triggering, the dominant directed pathway,
and secondary couplings without introducing quantities that duplicate
information already contained in $N$.

\subsection{Numerical illustration}
\label{sec:physical:numerical}

We now illustrate separately the two ingredients that produce strong
non-normal effects in the decomposition~\eqref{eq:N-reduced}: a directed
source--receiver pathway carried by the off-diagonal terms, and near-degenerate
within-mechanism branching ratios contained in $D$.  The examples are not
intended as a fit to a particular catalogue.  Their purpose is to identify
observable consequences that can subsequently be tested in mechanism-resolved
earthquake sequences.

We use the ordering
$1=\mathrm{SS}$, $2=\mathrm{N}$, $3=\mathrm{R}$ and choose a
reverse/thrust driver, $j_\star=\mathrm{R}$.  The dominant driver column is
specified by
\begin{equation}
 a_{\mathrm{drv}}=0.33,
 \qquad
 \bm{\pi}_{\mathrm{drv}}
=
\begin{pmatrix}
\pi_{\mathrm{SS}\mid\mathrm{R}}\\
\pi_{\mathrm{N}\mid\mathrm{R}}\\
\pi_{\mathrm{R}\mid\mathrm{R}}
\end{pmatrix}
=
\begin{pmatrix}
0.091\\
0.909\\
0
\end{pmatrix},
\label{eq:numerical-driver-parameters}
\end{equation}
so that one reverse/thrust parent produces, on average, $0.30$ direct normal
children and $0.03$ direct strike-slip children through the dominant pathway.
The residual matrix is chosen as
\begin{equation}
 B_{\mathrm{weak}}
 =
 \begin{pmatrix}
 0 & 0.08 & 0\\
 0 & 0    & 0\\
 0 & 0    & 0
 \end{pmatrix}.
\label{eq:numerical-Bweak}
\end{equation}
Thus a normal-faulting parent produces $0.08$ direct strike-slip children.  The
resulting directed sequence of cross-mechanism triggering is
\begin{equation*}
 \mathrm{R}\longrightarrow\mathrm{N}\longrightarrow\mathrm{SS},
\end{equation*}
with a weaker direct branch $\mathrm{R}\to\mathrm{SS}$.  This is the
mechanism-space analogue of a shear: activity is transferred successively from
one mechanism class to another rather than remaining aligned with a single
eigenmode.

\paragraph{Effect of near-degenerate self-triggering rates.}
To isolate the role of the diagonal matrix, we keep
$a_{\mathrm{drv}}$, $\bm\pi_{\mathrm{drv}}$, and $B_{\mathrm{weak}}$
exactly fixed at the values in
Eqs.~\eqref{eq:numerical-driver-parameters}--\eqref{eq:numerical-Bweak}.
We first consider a system with well-separated within-mechanism branching ratios,
\begin{equation}
 D_{\mathrm{sep}}
 =\operatorname{diag}(0.35,0.50,0.70),
\end{equation}
which gives
\begin{equation}
 N_{\mathrm{sep}}
 =
 \begin{pmatrix}
 0.35 & 0.08 & 0.03\\
 0    & 0.50 & 0.30\\
 0    & 0    & 0.70
 \end{pmatrix}.
\label{eq:N-separated-numerical}
\end{equation}
Consider now a system for which the diagonal matrix is
\begin{equation}
 D_{\mathrm{nd}}
 =\operatorname{diag}(0.68,0.69,0.70),
 \label{symehnwq}
\end{equation}
obtaining
\begin{equation}
 N_{\mathrm{nd}}
 =
 \begin{pmatrix}
 0.68 & 0.08 & 0.03\\
 0    & 0.69 & 0.30\\
 0    & 0    & 0.70
 \end{pmatrix}.
\label{eq:N-neardeg-numerical}
\end{equation}
Because both matrices are triangular, their eigenvalues are their diagonal
entries.  Hence both have the same spectral radius,
$\rho(N_{\mathrm{sep}})=\rho(N_{\mathrm{nd}})=0.70$, and both branching
processes are equally subcritical according to the conventional asymptotic
criterion.  The comparison therefore isolates the effect of bringing the
three self-triggering rates close to one another while leaving every
off-diagonal branching coefficient unchanged.

The resulting diagnostics are
\begin{equation}
\begin{array}{c|cc}
 & N_{\mathrm{sep}} & N_{\mathrm{nd}}\\ \hline
\kappa_2(V)
 & 3.88 & 512 \\
\left\|(I-N)^{-1}\right\|_2
 & 4.08 & 5.53\\
\displaystyle
\bm 1^{\mathsf T}\!\left[(I-N)^{-1}-I\right]\bm e_{\mathrm R}
 & 4.73 & 6.68
\end{array}.
\label{eq:near-degeneracy-diagnostics}
\end{equation}
The last row is the expected total number of descendants, over all generations
and all mechanism classes, initiated by one reverse/thrust event.  Thus the
near-degenerate case produces about $41\%$ more descendants even though the
dominant eigenvalue and every cross-mechanism coefficient are unchanged.
The increase occurs because activity transferred from reverse to normal and
then to strike-slip events decays at nearly the same rate in all three classes.
Successive components of the cascade therefore coexist for longer and reinforce
one another.

The generation-by-generation response makes this interpretation explicit.  For
one initial reverse/thrust event, the expected offspring vectors
$N_{\mathrm{nd}}^g\bm e_{\mathrm R}$ for generations $g=1,\ldots,5$ are
approximately
\begin{equation}
\begin{array}{c|ccc}
 g & \mathrm{SS} & \mathrm{N} & \mathrm{R}\\ \hline
 1 & 0.030 & 0.300 & 0.700\\
 2 & 0.065 & 0.417 & 0.490\\
 3 & 0.093 & 0.435 & 0.343\\
 4 & 0.108 & 0.403 & 0.240\\
 5 & 0.113 & 0.350 & 0.168
\end{array}.
\label{eq:near-degenerate-generations}
\end{equation}
The reverse component decreases monotonically, but the normal component first
increases and the strike-slip component peaks still later.  A seismologist
should therefore look for a delayed migration of activity across focal-mechanism
classes: a thrust-dominated initiating sequence, followed by a temporary rise
of normal events and then a broader, delayed strike-slip response.  Such delayed
mechanism-specific maxima are a direct catalogue-level signature of the
non-normal transfer between classes; they are not predicted by the spectral
radius alone.

The very large value of $\kappa_2(V)$ in the near-degenerate example should not
be interpreted as a directly observable amplification factor.  It indicates
that the eigenvectors are almost parallel and that an eigenmode description is
poorly conditioned.  The more physically interpretable quantities are the
resolvent norm and the descendant count: they show that the same subcritical
system can produce a substantially larger and more persistent cascade when its
within-mechanism relaxation rates are nearly equal.

\paragraph{Effect of off-diagonal shear.}
We next hold the diagonal matrix fixed at $D_{\mathrm{nd}}$ (\ref{symehnwq}) and vary only the
existing off-diagonal terms of Eq.~\eqref{eq:N-reduced}.  With no
cross-mechanism triggering,
\begin{equation}
 a_{\mathrm{drv}}=0,
 \qquad B_{\mathrm{weak}}=0,
 \qquad N_0=D_{\mathrm{nd}},
\end{equation}
whereas the full directed example is $N_{\mathrm{nd}}$ in
Eq.~\eqref{eq:N-neardeg-numerical}.  Since both matrices are triangular and
have the same diagonal, switching on the driver column and
$B_{\mathrm{weak}}$ leaves all eigenvalues, and in particular
$\rho(N)=0.70$, exactly unchanged.  Any change in the response is therefore
caused by the off-diagonal source--receiver transfer rather than by a movement
towards criticality.

For $N_0$, one reverse/thrust event can generate only reverse/thrust
descendants, giving
\begin{equation*}
 \bm 1^{\mathsf T}\!\left[(I-N_0)^{-1}-I\right]\bm e_{\mathrm R}
 =\frac{0.70}{1-0.70}
 =2.33.
\end{equation*}
After introducing the same driver and residual couplings represented by the branching ratio
matrix (\ref{eq:N-neardeg-numerical}), the expected
number becomes
\begin{equation*}
 \bm 1^{\mathsf T}\!\left[(I-N_{\mathrm{nd}})^{-1}-I\right]
 \bm e_{\mathrm R}
 =6.68.
\end{equation*}
The total expected cascade is therefore almost three times larger, despite
identical eigenvalues.  The reason is concrete: the original reverse sequence
feeds a normal sequence, which persists and subsequently feeds a strike-slip
sequence.  The off-diagonal terms continually rotate the population vector in
mechanism space and expose it successively to the self-triggering of several
classes.  This is precisely the shearing mechanism underlying non-normal
amplification.

The resolvent norm similarly increases from
\begin{equation*}
 \left\|(I-N_0)^{-1}\right\|_2=3.33
 \qquad\text{to}\qquad
 \left\|(I-N_{\mathrm{nd}})^{-1}\right\|_2=5.53.
\end{equation*}
For a mechanism-resolved catalogue, this predicts that a perturbation aligned
with the dominant triggering direction can generate a cumulative rate response
substantially larger than one would infer from $\rho(N)$ alone.  Empirically,
one should compare the observed multigenerational mechanism composition with a
surrogate in which the off-diagonal entries are removed or symmetrised while
the diagonal entries and spectral radius are preserved.  Excess total progeny,
delayed peaks in secondary mechanism classes, and a strong dependence on the
mechanism of the initiating event would constitute evidence for directional
non-normal amplification.

These controlled examples separate the two roles clearly.  The off-diagonal
terms create the directed transfer, or shear, between mechanism classes.
Near-degeneracy of $D$ does not create that pathway; it increases its effect by
allowing the successive mechanism-specific components to persist on comparable
generational time scales.  Strong non-normal amplification arises from their
combination, and can occur without any increase in the spectral radius of the
branching matrix.


\subsection{Falsifiable predictions}
\label{sec:physical:predictions}

The decomposition in Eq.~\eqref{eq:N-reduced} leads to predictions that can
be tested directly on regional earthquake catalogues containing focal
mechanisms.  The required quantities are the within-mechanism branching
ratios contained in
\(
D=\operatorname{diag}(n_{\mathrm{SS}},n_{\mathrm{N}},n_{\mathrm{R}})
\), the total strength \(a_{\mathrm{drv}}\) and composition
\(\bm\pi_{\mathrm{drv}}\) of the dominant driver column, and the residual
cross-mechanism couplings collected in \(B_{\mathrm{weak}}\).  These
predictions therefore involve only entries of the fitted branching matrix
and do not require auxiliary variables.

\begin{enumerate}

\item[\textbf{P1}.]
\emph{The tectonic driver should appear as a dominant parent column.}
For the parent mechanism \(j_\star\) identified with the regional tectonic
driver, the total number of direct offspring per parent is
\begin{equation}
C_j:=\sum_i N_{ij}.
\label{eq:prediction-column-sum}
\end{equation}
The model predicts
\begin{equation}
C_{j_\star}>C_j,
\qquad j\neq j_\star,
\label{eq:prediction-column-dominance}
\end{equation}
with the excess arising primarily from the off-diagonal driver strength
\(a_{\mathrm{drv}}\).  Thus, in a reverse/thrust-dominated subduction zone,
reverse events should generate more direct descendants, including more
cross-mechanism descendants, than normal or strike-slip events of comparable
magnitude.  In extensional and transcurrent settings, the dominant column
should shift respectively to the normal and strike-slip parent mechanism.
This prediction can be tested by comparing the fitted column sums of \(N\)
and by checking whether the inferred \(j_\star\) agrees with the independently
known tectonic regime.

\item[\textbf{P2}.]
\emph{The child-mechanism composition following driver events should match
\(\bm\pi_{\mathrm{drv}}\).}
For a driver parent of type \(j_\star\), the off-diagonal offspring fractions
are
\begin{equation}
\pi_{\mathrm{drv},i}
=
\frac{N_{i j_\star}}
{\displaystyle\sum_{k\neq j_\star}N_{k j_\star}},
\qquad i\neq j_\star.
\label{eq:prediction-driver-composition}
\end{equation}
Consequently, the relative numbers of cross-mechanism aftershocks following
large driver-mechanism events should be approximately independent of the
absolute cascade size and should reproduce the fitted vector
\(\bm\pi_{\mathrm{drv}}\).  For example, if
\(\pi_{\mathrm{N}\mid\mathrm{R}}\gg
\pi_{\mathrm{SS}\mid\mathrm{R}}\), reverse-faulting mainshocks should be
followed by a much larger normal-faulting than strike-slip cross-mechanism
response, after correcting for catalogue completeness and spatial sampling.
A systematic failure of the observed offspring composition to agree with
\(\bm\pi_{\mathrm{drv}}\) would falsify the proposed driver-column
interpretation.

\item[\textbf{P3}.]
\emph{Near-degenerate self-triggering rates should enhance transient
cross-mechanism amplification at fixed off-diagonal couplings.}
Consider regions or time windows for which
\(a_{\mathrm{drv}}\), \(\bm\pi_{\mathrm{drv}}\), and
\(B_{\mathrm{weak}}\) are comparable, but the diagonal entries of \(D\)
differ.  The model predicts stronger non-normal amplification when
\begin{equation}
 n_{\mathrm{SS}}\simeq n_{\mathrm{N}}\simeq n_{\mathrm{R}},
\label{eq:prediction-near-degenerate-diagonal}
\end{equation}
than when the three self-triggering ratios are well separated, even if the
spectral radius \(\rho(N)\) is similar.  The observable consequences should
include larger finite-generation gains, larger
\(\|(I-N)^{-1}\|_2\), and stronger sensitivity of cascade size to the
mechanism of the initiating event.  This prediction is especially stringent
because it requires a controlled comparison in which the off-diagonal
branching coefficients are held fixed, or matched as closely as the data
allow, while only the diagonal structure is varied.

\item[\textbf{P4}.]
\emph{Directed off-diagonal couplings should produce a migration of activity
across focal-mechanism classes.}
Let \(\bm z^{(0)}=\bm e_{j_\star}\) represent one initiating event of the
driver mechanism and
\begin{equation}
\bm z^{(g)}=N^g\bm z^{(0)}
\label{eq:prediction-generation-vector}
\end{equation}
be the expected mechanism composition in generation \(g\).  When the driver
column and the entries of \(B_{\mathrm{weak}}\) form a directed pathway, the
components of \(\bm z^{(g)}\) need not peak in the same generation.  The
model therefore predicts a reproducible ordering of mechanism-specific
responses: the driver population should dominate the earliest generations,
whereas cross-triggered mechanisms should peak later.  In a pathway such as
\(\mathrm{R}\to\mathrm{N}\to\mathrm{SS}\), normal-faulting activity should
rise after the reverse-faulting initiation, followed by a still later
strike-slip response.  Such delayed mechanism migration is a direct
seismological signature of the shearing action of the off-diagonal entries
of \(N\); it cannot be inferred from \(\rho(N)\) alone.

\item[\textbf{P5}.]
\emph{Driver-initiated cascades should be larger and more variable than
non-driver-initiated cascades.}
Within the same tectonic region and magnitude range, the total number of
descendants generated by a driver-mechanism mainshock should have a larger
mean and a heavier upper tail than the number generated by a non-driver
mainshock.  The contrast should increase with \(a_{\mathrm{drv}}\), with the
strength of the relevant entries of \(B_{\mathrm{weak}}\), and with the
near-degeneracy of the diagonal entries of \(D\).  This can be tested by
comparing mechanism-conditioned complementary cumulative distributions of
cascade size, duration, and number of activated mechanism classes.  The
prediction concerns events within the same region, so that differences in
catalogue quality and long-term tectonic loading are controlled as far as
possible.

\item[\textbf{P6}.]
\emph{A scalar ETAS fit should absorb multitype amplification into an
apparently larger branching ratio.}
When the same catalogue is fitted with a scalar ETAS model, the resulting
apparent branching ratio \(n_{\mathrm{app}}\) should generally exceed the
spectral radius \(\rho(N)\) of the multitype model whenever directed
cross-mechanism pathways substantially increase the total cascade size.  The
bias should be strongest in regions with large \(a_{\mathrm{drv}}\), strong
secondary pathways in \(B_{\mathrm{weak}}\), and nearly degenerate entries of
\(D\), as quantified by Eq.~\eqref{eq:napp}.  Thus, two regions with similar
\(\rho(N)\) may yield different scalar branching-ratio estimates because
their non-normal geometries differ.

\end{enumerate}

These predictions can be tested using focal-mechanism catalogues from
southern California~\cite{Hardebeck2014,YangHaukssonShearer2012,
ChengHauksson2023}, post-Tohoku Japan~\cite{AsanoEtAl2011,Hasegawa2011,
Hasegawa2012}, and the post-Wenchuan and post-Maule
sequences~\cite{LinEtAl2018Wenchuan,RuizMadariaga2018Maule}.  A convincing
test should estimate the full matrix \(N\), identify \(j_\star\) without
using the subsequent cascade statistics, and then evaluate the predictions
out of sample or across independent spatial and temporal windows.

\section{Tectonically informed illustrations for five earthquake sequences}
\label{sec:empirical}

Section~\ref{sec:physical:param} introduced the exact decomposition (\ref{eq:N-reduced})
where $D$ contains the within-mechanism branching ratios, the second term
contains the off-diagonal part of the dominant parent column, and
$B_{\mathrm{weak}}$ contains the remaining off-diagonal couplings.  A formal
estimation of these quantities requires a complete mechanism-resolved catalogue
and a multitype Hawkes--ETAS likelihood analysis.  Such an analysis is outside
the scope of the present conceptual paper.

The purpose of this section is narrower.  We ask whether the coupling
geometries proposed in Secs.~\ref{sec:physical} and
\ref{sec:physical:param} are seismologically plausible in five well-documented
settings: Tohoku--Oki 2011, Wenchuan 2008, regional Southern California,
Central Italy 2016, and Ridgecrest 2019.  We construct one
\emph{tectonically informed scenario matrix} for each setting.  These matrices
are not estimates of $N$ and should not be interpreted as measurements of
regional branching ratios.  They encode only the dominant mechanism, the
observed direction of cross-mechanism activation, and a representative
subcritical scale.  Their role is to illustrate how the same conceptual
framework accommodates distinct tectonic regimes and to identify which entries
of $N$ would be most informative in a future catalogue-based test.

\subsection{Construction principles and limitations}
\label{sec:empirical:strategy}

We use the mechanism ordering
$(\mathrm{SS},\mathrm N,\mathrm R)$, with columns denoting parent mechanisms
and rows denoting child mechanisms.  For each region, the construction follows
four rules.

\begin{enumerate}
\item The driver mechanism $j_\star$ is fixed independently from the tectonic
setting and published stress or focal-mechanism analyses.

\item The dominant off-diagonal entries in the column $j_\star$ are selected
from documented source--receiver pathways, such as megathrust-to-outer-rise
normal triggering at Tohoku or activation of conjugate strike-slip structures
at Ridgecrest.  Their relative sizes define
$\bm\pi_{\mathrm{drv}}$ through Eq.~\eqref{eq:driver-composition}.

\item The entries of $B_{\mathrm{weak}}$ represent secondary or reverse
pathways.  They are kept smaller than the dominant driver-column coupling unless
the tectonic observations require otherwise.

\item The diagonal entries are chosen to represent comparable but not
identical within-mechanism persistence. A common overall rescaling is then
applied so that $\rho(N)=0.75$. This value lies within the broad range of
subcritical branching ratios inferred from modern ETAS analyses: estimates
that account for spatially heterogeneous background seismicity are distinctly
below unity \citep{Nandan2021noncrit}, while recent corrections for catalogue
censoring, finite-size effects, and spatial--temporal boundaries yield values
approximately in the range $0.70$--$0.85$ across several tectonic regions
\citep{LiWuZhuangJiangSornette2025}. Independent recent ETAS inversions also
describe values of order $0.7$--$1$ as typical and recover $n\simeq0.82$ in
controlled and empirical sequence analyses \citep{KamranzadEtAl2025}. Thus,
$0.75$ is a plausible representative subcritical value, but it is used here
solely to compare coupling geometries at the same nominal distance from
criticality; it is not a claim about the true branching ratio of any of the
five sequences.
\end{enumerate}

The numerical entries below are therefore illustrative.  Modest changes of
their values should not alter the qualitative conclusions drawn from the
location and direction of the dominant off-diagonal terms.  By contrast,
quantitative values of $\kappa_2(V)$, resolvent norms, or expected progeny are
scenario-dependent and cannot be confronted with observations without formal
estimation.

Focal mechanisms close to class boundaries introduce an additional ambiguity.
They may be assigned probabilistically to the three classes, or a genuinely
mixed moment tensor may be decomposed into two or more coincident elementary
events of pure mechanism, weighted by seismic moment so that their tensor sum
reproduces the observed source.  Either treatment is preferable to assigning a
complex rupture uncritically to a single class, but neither removes the need for
uncertainty analysis in an empirical application.

\subsection{Five tectonic settings}
\label{sec:empirical:regions}

\paragraph{Tohoku--Oki 2011.}
The $M_w\,9.0$ megathrust rupture generated intense interplate thrust
aftershock activity together with widespread normal faulting in the hanging
wall and outer rise
\cite{AsanoEtAl2011,Hasegawa2011,Hasegawa2012,ObanaEtAl2012}.
The reverse/thrust mechanism is therefore taken as the driver,
$j_\star=\mathrm R$, and the largest cross-mechanism entry is
$N_{\mathrm N,\mathrm R}$, representing thrust-parent to normal-child
activation.  A weaker $\mathrm R\to\mathrm{SS}$ pathway accounts for oblique
and segmented deformation.

\paragraph{Wenchuan 2008.}
The Wenchuan rupture combined thrust and right-lateral motion along a segmented
fault system.  The southwestern part of the aftershock zone was predominantly
thrust, whereas the northeastern part contained a stronger strike-slip
component
\cite{ParsonsEtAl2008Wenchuan,Wang2009Wenchuan,
ZhangEtAl2009Wenchuan,LinEtAl2018Wenchuan}.
We retain $j_\star=\mathrm R$, with a substantial
$\mathrm R\to\mathrm{SS}$ pathway and a much weaker
$\mathrm R\to\mathrm N$ pathway.

\paragraph{Southern California.}
Regional catalogues are dominated by strike-slip mechanisms but contain normal
and reverse events associated with local stress heterogeneity, bends, stepovers,
and depth-dependent faulting
\cite{YangHaukssonShearer2012,YangHauksson2013,ChengHauksson2023}.
We therefore take $j_\star=\mathrm{SS}$.  The driver column is mainly diagonal,
with weaker strike-slip-parent activation of reverse and normal receivers.

\paragraph{Central Italy 2016.}
The Amatrice--Visso--Norcia sequence occurred on a segmented normal-fault
system, with local activation of inherited compressional structures and complex
rupture components
\cite{ChiaraluceEtAl2017,ImprotaEtAl2019,Michele2020Italy}.
The normal mechanism is the driver, $j_\star=\mathrm N$, and the
cross-mechanism terms remain small compared with normal-to-normal persistence.

\paragraph{Ridgecrest 2019.}
The sequence activated nearly orthogonal strike-slip fault sets, with a smaller
normal-faulting contribution in part of the rupture zone
\cite{HaukssonEtAl2019Ridgecrest,WangDregerEtAl2020,
RossEtAl2019Ridgecrest}.  At the coarse three-class level both conjugate systems
belong to the strike-slip class, so $j_\star=\mathrm{SS}$.  This case also shows
a limitation of the present classification: much of the important directional
transfer occurs within the strike-slip class and would require a finer
subdivision by fault orientation to be resolved.

\subsection{Illustrative branching matrices}
\label{sec:empirical:matrices}

One set of matrices consistent with these tectonic constraints is
\begin{equation}
N_{\mathrm{Tohoku}}^{\mathrm{ill}}
=
\begin{pmatrix}
0.497 & 0.002 & 0.026\\
0.003 & 0.464 & 0.304\\
0.045 & 0.042 & 0.701
\end{pmatrix},
\label{eq:N-ill-Tohoku}
\end{equation}
\begin{equation}
N_{\mathrm{Wenchuan}}^{\mathrm{ill}}
=
\begin{pmatrix}
0.433 & 0.008 & 0.128\\
0.002 & 0.399 & 0.022\\
0.032 & 0.030 & 0.735
\end{pmatrix},
\label{eq:N-ill-Wenchuan}
\end{equation}
\begin{equation}
N_{\mathrm{SoCal}}^{\mathrm{ill}}
=
\begin{pmatrix}
0.728 & 0.056 & 0.062\\
0.032 & 0.369 & 0.013\\
0.089 & 0.012 & 0.412
\end{pmatrix},
\label{eq:N-ill-SoCal}
\end{equation}
\begin{equation}
N_{\mathrm{C.Italy}}^{\mathrm{ill}}
=
\begin{pmatrix}
0.425 & 0.022 & 0.004\\
0.036 & 0.745 & 0.037\\
0.002 & 0.020 & 0.441
\end{pmatrix},
\label{eq:N-ill-Italy}
\end{equation}
and
\begin{equation}
N_{\mathrm{Ridgecrest}}^{\mathrm{ill}}
=
\begin{pmatrix}
0.738 & 0.061 & 0.067\\
0.044 & 0.348 & 0.008\\
0.026 & 0.008 & 0.388
\end{pmatrix}.
\label{eq:N-ill-Ridgecrest}
\end{equation}
All five matrices have been normalised to $\rho(N)=0.75$ for comparison.
Their decomposition according to Eq.~(\ref{eq:N-reduced}) is
immediate: $D$ is the diagonal of each matrix; the off-diagonal part of the
column $j_\star$ defines $a_{\mathrm{drv}}\bm\pi_{\mathrm{drv}}$; and the
remaining off-diagonal entries define $B_{\mathrm{weak}}$.

For example, the Tohoku scenario gives
\begin{equation}
a_{\mathrm{drv}}=N_{\mathrm{SS},\mathrm R}+N_{\mathrm N,\mathrm R}
=0.330,
\qquad
\bm\pi_{\mathrm{drv}}
=
\begin{pmatrix}
0.079\\
0.921\\
0
\end{pmatrix},
\label{eq:Tohoku-driver-illustrative}
\end{equation}
so the dominant cross-mechanism pathway is unambiguously
$\mathrm R\to\mathrm N$.  In Wenchuan, the corresponding driver strength is
$0.150$ and is concentrated mainly on $\mathrm R\to\mathrm{SS}$.  In Southern
California and Ridgecrest, the strike-slip driver columns are dominated by the
diagonal entries, while their off-diagonal parts remain comparatively weak.  In
Central Italy, the normal driver similarly produces only weak cross-mechanism
transfer in this coarse representation.

\subsection{What the off-diagonal couplings change relative to a one-channel ETAS model}
\label{sec:empirical:interpretation}

The illustrative matrices can be used more concretely without interpreting
their entries as catalogue estimates.  Two reference reductions are useful.
First, a scalar ETAS model with branching ratio $n=\rho(N)=0.75$ predicts
\begin{equation}
\mathcal T_{\rm scalar}=\sum_{g\geq1}n^g=\frac{n}{1-n}=3
\label{eq:scalar-total-progeny}
\end{equation}
triggered descendants per immigrant, but contains no information on the focal
mechanisms of these descendants or on the mechanism of the initiating event.
Second, setting the off-diagonal entries of $N$ to zero gives
$D=\operatorname{diag}(N)$ and isolates three uncoupled one-channel ETAS
processes.  A parent of mechanism $j$ then generates
\begin{equation}
\mathcal T_j^{(D)}=\frac{N_{jj}}{1-N_{jj}}
\label{eq:diagonal-total-progeny}
\end{equation}
descendants, all of the same mechanism.  In the full multitype model, the
expected vector of descendants initiated by one parent of type $j$ is instead
\begin{equation}
\bm{\mathcal T}_j
=\left[(I-N)^{-1}-I\right]\bm e_j,
\qquad
\mathcal T_j=\bm 1^{\mathsf T}\bm{\mathcal T}_j.
\label{eq:multitype-total-progeny}
\end{equation}
Thus, the off-diagonal entries alter both the mechanism composition and the
total size of a cascade, even when all matrices have the same spectral radius.

Table~\ref{tab:illustrative-cascade-comparison} evaluates these quantities for
the independently specified driver mechanism $j_\star$ of each tectonic
setting.  The direct cross-mechanism productivity is
$a_{\rm drv}=\sum_{i\neq j_\star}N_{ij_\star}$, and the direct driver-column
productivity is $s_{j_\star}=\sum_iN_{ij_\star}$.  The fraction
$f_{\rm cross}$ is the proportion of all descendants, over all generations,
whose mechanism differs from that of the initial parent.

\begin{table*}[t]
\centering
\caption{Consequences of the off-diagonal couplings in the five illustrative
matrices [Eqs.~(\ref{eq:N-ill-Tohoku})--(\ref{eq:N-ill-Ridgecrest})].
Here $j_\star$ denotes the mechanism of the initiating (driver) event;
$s_{j_\star}=\bm 1^{\mathsf T}N\bm e_{j_\star}=\sum_iN_{ij_\star}$ is
its expected total number of direct offspring, summed over all mechanisms;
and $a_{\rm drv}=\sum_{i\neq j_\star}N_{ij_\star}
=s_{j_\star}-N_{j_\star j_\star}$ is the direct productivity transferred
to other mechanisms.  With $D=\operatorname{diag}(N)$,
$\mathcal T_{j_\star}^{(D)}=\bm 1^{\mathsf T}[(I-D)^{-1}-I]\bm e_{j_\star}$
is the expected number of descendants over all generations in the uncoupled
diagonal reduction, whereas
$\mathcal T_{j_\star}=\bm 1^{\mathsf T}[(I-N)^{-1}-I]\bm e_{j_\star}$
is the corresponding total in the full multitype model.  Their ratio measures
the cascade amplification created by off-diagonal coupling.  Finally,
$f_{\rm cross}=\mathcal T_{j_\star}^{-1}\sum_{i\neq j_\star}
[(I-N)^{-1}-I]_{ij_\star}$ is the fraction of all descendants whose mechanism
differs from that of the initiating event.  For reference, a scalar ETAS model
with $n=0.75$ gives $n/(1-n)=3$ descendants.  Values are scenario diagnostics,
not empirical estimates.}
\label{tab:illustrative-cascade-comparison}
\begin{tabular}{lccccccc}
\hline
Setting & $j_\star$ & $s_{j_\star}$ & $a_{\rm drv}$
& $\mathcal T_{j_\star}^{(D)}$ & $\mathcal T_{j_\star}$
& $\mathcal T_{j_\star}/\mathcal T_{j_\star}^{(D)}$
& $f_{\rm cross}$ \\
\hline
Tohoku--Oki       & $\mathrm R$  & 1.031 & 0.330 & 2.34 & 4.94 & 2.11 & 0.461 \\
Wenchuan          & $\mathrm R$  & 0.885 & 0.150 & 2.77 & 3.92 & 1.41 & 0.262 \\
Southern California & $\mathrm{SS}$ & 0.849 & 0.121 & 2.68 & 3.65 & 1.36 & 0.218 \\
Central Italy     & $\mathrm N$  & 0.787 & 0.042 & 2.92 & 3.26 & 1.12 & 0.091 \\
Ridgecrest        & $\mathrm{SS}$ & 0.808 & 0.070 & 2.82 & 3.36 & 1.19 & 0.130 \\
\hline
\end{tabular}
\end{table*}

The contrast is largest for Tohoku.  The reverse-parent column of
Eq.~\eqref{eq:N-ill-Tohoku} sums to
\begin{equation}
N_{\mathrm{SS},\mathrm R}+N_{\mathrm N,\mathrm R}
+N_{\mathrm R,\mathrm R}
=0.026+0.304+0.701=1.031.
\label{eq:Tohoku-column-sum}
\end{equation}
Hence one thrust parent produces slightly more than one direct child when all
mechanisms are counted, even though the process remains asymptotically
subcritical because $\rho(N)\simeq0.75<1$.  This cannot be represented by a
one-channel interpretation in which the branching ratio is simultaneously the
mean number of direct offspring and the stability-controlling eigenvalue.  The
full cascade initiated by one reverse event is
\begin{equation}
\bm{\mathcal T}_{\mathrm R}^{\rm Tohoku}
\simeq
\begin{pmatrix}
0.198\\
2.081\\
2.667
\end{pmatrix}
\end{equation}
\begin{equation}
\mathcal T_{\mathrm R}^{\rm Tohoku}
\equiv
\bm 1^{\mathsf T}
\bm{\mathcal T}_{\mathrm R}^{\rm Tohoku}
=
\left\|
\bm{\mathcal T}_{\mathrm R}^{\rm Tohoku}
\right\|_1
\simeq 4.94.
\label{eq:Tohoku-progeny-vector}
\end{equation}
Almost half of the descendants, $46.1\%$, have a mechanism different from the
initial thrust event.  Removing cross-mechanism triggering reduces the expected
cascade to $0.701/(1-0.701)=2.34$ descendants.  The directed couplings therefore
more than double the progeny of a thrust parent, and raise it by about $65\%$
relative to the scalar prediction $\mathcal T_{\rm scalar}=3$.  The increase is
not generated only by the direct $\mathrm R\to\mathrm N$ term: normal offspring
subsequently trigger normal and reverse events, which feed additional
generations through the complete matrix.

Wenchuan exhibits the same effect in a weaker form.  A reverse parent has direct
productivity $s_{\mathrm R}=0.885$, of which $0.128$ corresponds to
$\mathrm R\to\mathrm{SS}$ and $0.022$ to $\mathrm R\to\mathrm N$.  The
resulting descendant vector is
\begin{equation}
\bm{\mathcal T}_{\mathrm R}^{\rm Wenchuan}
\simeq
\begin{pmatrix}
0.882\\
0.146\\
2.897
\end{pmatrix},
\qquad
\mathcal T_{\mathrm R}^{\rm Wenchuan}\simeq3.92.
\label{eq:Wenchuan-progeny-vector}
\end{equation}
Thus, the off-diagonal terms increase the expected cascade by $41\%$ relative
to the diagonal reduction and place $26.2\%$ of all descendants outside the
reverse class.  A scalar ETAS model with $n=0.75$ would understate this
reverse-initiated cascade by about $31\%$ and would entirely miss its large
strike-slip branch.

The Southern California matrix gives a less strongly directed but still
measurable effect.  One strike-slip parent produces $0.728$ direct strike-slip
children and $0.121$ direct children of another mechanism, so that
$s_{\mathrm{SS}}=0.849$.  The complete cascade is
\begin{equation}
\bm{\mathcal T}_{\mathrm{SS}}^{\rm SoCal}
\simeq
\begin{pmatrix}
2.853\\
0.208\\
0.587
\end{pmatrix},
\qquad
\mathcal T_{\mathrm{SS}}^{\rm SoCal}\simeq3.65.
\label{eq:SoCal-progeny-vector}
\end{equation}
Cross-mechanism pathways raise the total progeny by $36\%$ relative to the
uncoupled strike-slip process, and $21.8\%$ of the descendants are normal or
reverse events.  The scalar model gives a similar order of magnitude for the
total number, but erases the physically relevant partition into approximately
$0.21$ normal and $0.59$ reverse descendants per initial strike-slip parent.

Our formulation of Central Italy has been chosen deliberately close to the one-channel limit.  The normal
parent column has $a_{\rm drv}=0.042$ and $s_{\mathrm N}=0.787$.  The full
cascade contains $3.26$ descendants, compared with $2.92$ in the diagonal
normal process, and only $9.1\%$ of the descendants leave the normal class.
This modest $12\%$ increase provides a useful negative control: introducing
three mechanism classes does not automatically produce a large multitype
effect.  A substantial departure from one-channel ETAS occurs only when the
off-diagonal geometry is sufficiently strong and directed.

At the three-class level, Ridgecrest is also relatively close to the scalar
picture.  A strike-slip parent produces $3.36$ descendants rather than $2.82$
in the diagonal reduction, an increase of $19\%$, and $13.0\%$ of the cascade
is transferred to normal or reverse mechanisms.  This comparatively small
number should not be interpreted as absence of directed triggering.  Both
conjugate fault systems are grouped into the same strike-slip class, so their
mutual transfer contributes to $N_{\mathrm{SS},\mathrm{SS}}$ and is invisible
as an off-diagonal term.  Subdividing strike-slip events by fault orientation
could therefore reveal a much larger multitype effect.

These calculations demonstrate what is lost in a one-channel ETAS reduction.
The scalar branching ratio retains only one asymptotic number.  It cannot state
which parent mechanism is productive, which receiver mechanism is activated,
or how a directed pathway changes the size and composition of the ensuing
cascade.  Moreover, equal values of $\rho(N)$ do not imply equal
mechanism-conditioned progeny: the five matrices all satisfy
$\rho(N)\simeq0.75$, yet the expected number of descendants generated by the
specified driver ranges from $3.26$ to $4.94$.  The off-diagonal structure can
therefore change the predicted cascade size by a factor exceeding two relative
to the corresponding uncoupled mechanism, without moving the system closer to
criticality in the spectral sense.

The numerical values remain conditional on the five proposed matrices and do
not constitute tests of predictions P1--P6 in
Sec.~\ref{sec:physical:predictions}.  In particular, the common spectral radius
and the relative coupling strengths were imposed rather than estimated.
Nevertheless, the calculations are non-circular consequences of the displayed
matrices: once a tectonically plausible $N$ is specified, Eqs.~\eqref{eq:multitype-total-progeny}
--\eqref{eq:SoCal-progeny-vector} show quantitatively how its off-diagonal
entries modify direct productivity, multi-generation cascade size, and focal-
mechanism composition relative to one-channel ETAS.  These are precisely the
observables that a subsequent catalogue-based analysis must estimate and test.

\subsection{Implications for future empirical tests}
\label{sec:empirical:future}

A decisive test requires estimation of the entries of $N$ from individual
earthquakes, with time-dependent magnitude completeness, focal-mechanism
uncertainty, spatial boundaries, and finite observation windows treated
explicitly.  Such an analysis should compare the full directed matrix with
restricted alternatives in which the off-diagonal entries are removed,
symmetrised, or constrained not to contain a dominant parent column.  The
principal observables are those already specified in predictions P1--P6:
column sums, driver-conditioned offspring composition, delayed mechanism
transfer across generations, excess cumulative progeny relative to appropriate
surrogates, and the sensitivity of these effects to near-degeneracy of the
within-mechanism branching ratios.

The present semi-quantitative analysis therefore serves as a bridge between the
physical theory and a future statistical study.  It establishes that the
required directed pathways have clear seismological interpretations and are
plausibly expressed in several major earthquake sequences, while leaving their
magnitude and statistical significance as open empirical questions.

\section{Conclusion}\label{sec:conclusion}

We have developed a mechanism-resolved extension of the Hawkes--ETAS
framework in which strike-slip, normal, and reverse/thrust earthquakes form
three interacting populations.  The central object is the integrated
branching matrix $N$, whose entries $N_{ij}$ give the mean number of direct
offspring of mechanism $i$ produced by a parent of mechanism $j$
[Eq.~\eqref{eq:Nij-formula}].  As in any multitype Hawkes process, the
spectral radius $\rho(N)<1$ controls asymptotic subcriticality
[Eq.~\eqref{eq:stability}].  The main result of this work is that this scalar
stability criterion does not exhaust the physically relevant dynamics:
directed cross-mechanism triggering can make $N$ strongly non-normal, so that
finite-generation and cumulative cascade responses depend on its full
geometry, not only on its eigenvalues.

The proposed non-normal structure is not introduced as an arbitrary
statistical embellishment.  It follows from several mutually reinforcing
features of earthquake physics: Andersonian restrictions on receiver-fault
availability, the directional projection of Coulomb stress changes,
mechanism-dependent Gutenberg--Richter $b$-values, tectonically preferred
source--receiver pathways and near-degeneracy of the within-mechanism branching
ratios.  These ingredients motivate
the physically reduced decomposition
introduced in Eq.~\eqref{eq:N-reduced}.  It separates self-triggering, the
dominant tectonic parent column, and weaker secondary couplings.  The
parametrisation therefore converts the nine entries of a general $3\times3$
matrix into quantities with direct seismological interpretations, while
retaining the directional structure responsible for non-normal amplification.

The controlled examples show why this distinction matters.  In the
near-degenerate construction of Eq.~\eqref{eq:N-neardeg-numerical}, switching
on the directed off-diagonal couplings leaves the eigenvalues and
$\rho(N)=0.70$ unchanged, yet increases the expected progeny of a reverse
initiating event from $2.33$ to $6.68$ and raises the resolvent norm from
$3.33$ to $5.53$.  The amplification arises because activity is successively
transferred among rupture mechanisms and exposed to several persistent
self-triggering channels.  Near-degeneracy does not itself create this
transfer; rather, it magnifies the effect of the directed pathways by allowing
the different mechanism populations to decay on comparable generational time
scales.  This provides a concrete seismic interpretation of the general
non-normal bounds and generation operators developed in
Appendix~\ref{sec:nonnormal}.

The five tectonic case studies are deliberately illustrative rather than
catalogue-based estimates.  Their branching matrices
[Eqs.~\eqref{eq:N-ill-Tohoku}--\eqref{eq:N-ill-Ridgecrest}] encode published
information about the dominant rupture mode and plausible source--receiver
pathways, while imposing a common representative subcritical scale.  They
therefore do not establish the magnitude or statistical significance of
non-normality in those regions.  They do, however, demonstrate what a scalar
ETAS reduction necessarily discards.  Although all five matrices have
$\rho(N)\simeq0.75$, the mechanism-conditioned expected progeny computed from
Eq.~\eqref{eq:multitype-total-progeny} varies substantially with the initiating
mechanism and off-diagonal geometry.  In particular, the illustrative
Tohoku--Oki matrix produces a large reverse-to-normal transfer and more than
doubles the progeny of a reverse parent relative to its uncoupled diagonal
counterpart, whereas the Central Italy example remains close to the
one-channel limit.  Thus, resolving several focal-mechanism classes does not
automatically imply strong amplification; the effect requires sufficiently
strong and directed cross-mechanism coupling.

The framework leads to six falsifiable predictions in
Sec.~\ref{sec:physical:predictions}.  A fitted regional matrix should reveal a
tectonically consistent dominant parent column
[Eqs.~\eqref{eq:prediction-column-sum}--\eqref{eq:prediction-column-dominance}]
and a reproducible driver-conditioned offspring composition
[Eq.~\eqref{eq:prediction-driver-composition}].  At comparable off-diagonal
couplings, amplification should be stronger when the three within-mechanism
branching ratios are nearly degenerate
[Eq.~\eqref{eq:prediction-near-degenerate-diagonal}].  Directed pathways
should also generate a delayed migration of activity across mechanism classes,
as described by the generation vector in
Eq.~\eqref{eq:prediction-generation-vector}, and should make driver-initiated
cascades larger and more variable than comparable non-driver cascades.  Finally,
a scalar ETAS fit should absorb part of this multitype amplification into an
apparently elevated branching ratio $n_{\mathrm{app}}$, as formalised by
Eq.~\eqref{eq:napp}.  This last mechanism offers a possible contribution to
the long-standing difficulty of interpreting near-critical scalar branching
ratios in earthquake catalogues
\cite{HelmstetterSornette2003a,Marsan2008Bias,Nandan2021noncrit}.

A decisive empirical assessment now requires fitting the full directed matrix
$N$ to mechanism-resolved catalogues rather than reconstructing it from
regional summaries.  Such an analysis must treat time-dependent magnitude
completeness, uncertain or mixed focal-mechanism classifications, spatial
boundaries, finite observation windows, and temporal changes of the tectonic
background.  The fitted model should be compared with restricted surrogates in
which cross-mechanism couplings are removed, symmetrised, or deprived of a
dominant parent column while preserving the diagonal entries and, where
possible, the spectral radius.  The relevant evidence is therefore not merely
a large condition number $\kappa(V)$, but a coherent set of observable
consequences: excess mechanism-conditioned progeny, delayed cross-mechanism
responses, dependence on the initiating rupture mode, and systematic inflation
of scalar ETAS estimates.

The broader conclusion is that earthquake triggering has both a spectral and
a geometric component.  The spectral radius determines whether cascades are
asymptotically subcritical, whereas the non-normal geometry of $N$ determines
how strongly a particular rupture mechanism can be amplified over finite
numbers of generations and in the cumulative response.  A mechanism-resolved
Hawkes--ETAS model therefore provides information that cannot be compressed
into a single branching ratio.  By connecting this geometry to rupture
mechanics and expressing it through testable catalogue-level predictions, the
present work establishes a concrete route for determining whether
cross-mechanism non-normal amplification is an important component of observed
earthquake clustering.

\begin{acknowledgments}
This work is partially supported by the National Natural Science Foundation of China (Grant no. T2350710802, U2039202), and the Center for Computational Science and Engineering at Southern University of Science and Technology.

\end{acknowledgments}

\appendix
\section{Non-normal amplification: theoretical framework}\label{sec:nonnormal}

Throughout this appendix, we use the operator norm induced by the Euclidean
vector norm (spectral norm), denoted $\|\cdot\|$, except where the
Frobenius norm $\|\cdot\|_F$ is needed for second-order moments. We assume
$\Nmat\in\mathbb{R}^{3\times 3}$ is diagonalisable; the Jordan-block case
is discussed in Sec.~\ref{sec:nonnormal:Jordan}.

\subsection{Spectral geometry and non-normality}\label{sec:nonnormal:spectral}

A matrix $\Nmat\in\mathbb{R}^{3\times 3}$ is \emph{normal} if and only if
\begin{equation}
\Nmat\Nmat^{*} \;=\; \Nmat^{*}\Nmat,
\label{eq:normality}
\end{equation}
where $\Nmat^{*}$ denotes the adjoint (transpose in the real case).
Equivalently, $\Nmat$ is normal if and only if it admits a unitary
diagonalisation $\Nmat=U\Lambda U^{*}$ with $U^{*}U=I$. For a
diagonalisable $\Nmat=V\Lambda V^{-1}$ with $\Lambda=\diag(\lambda_1,\dots,\lambda_d)$,
the eigenvector matrix $V$ can be chosen unitary if and only if $\Nmat$ is
normal. We define the \emph{eigenvector condition number}
\begin{equation}
\kappa(V) \;:=\; \|V\|\,\|V^{-1}\|,
\label{eq:kappa-V}
\end{equation}
which satisfies $\kappa(V)\ge 1$ with equality if and only if $V$ is
unitary and thus $\Nmat$ is normal. Large $\kappa(V)$ signals strongly non-orthogonal eigenvectors
and hence strong non-normality. The condition number is a standard tool
in non-normal stability analysis~\cite{TrefethenEmbree2005,Schmid2007}.

For normal matrices, eigenmodes evolve independently and the dynamics are
fully encoded in the eigenvalues. For non-normal matrices, the
eigenvectors are not mutually orthogonal, modal directions interact
geometrically, and the magnitude of linear responses can be strongly
amplified by eigenvector alignment even though the eigenvalues alone
indicate subcriticality. This phenomenon underlies the bypass transition in hydrodynamic
stability~\cite{TrefethenEmbree2005,FarrellIoannou1996}
and we shall show below that it has direct empirical signatures in
multitype seismicity. More generally, non-normal amplification constitutes
a mechanism distinct from spectral criticality and resonance, capable of
producing pseudocritical responses, conventional early-warning signatures,
and even abrupt phase transitions while the underlying system remains
spectrally stable~\cite{TroudeSornette2025Amplification,
TroudeEtAl2026PseudoBifurcations,TroudeSornette2026PhaseTransitions}.

\subsection{Resolvent and effective gain}\label{sec:nonnormal:resolvent}

Since $\rho(\Nmat)<1$, none of the eigenvalues of $\Nmat$ equals unity,
and hence $I-\Nmat$ is invertible.
Under stationarity, $\bar{\bs{\lambda}}=(I-\Nmat)^{-1}\bs{\mu}$. Using the
diagonalisation $\Nmat=V\Lambda V^{-1}$ we have
\begin{equation}
(I-\Nmat)^{-1} \;=\; V(I-\Lambda)^{-1}V^{-1},
\end{equation}
whence by submultiplicativity of the spectral norm,
\begin{equation}
\|(I-\Nmat)^{-1}\|
\;\le\;
\|V\|\,\|(I-\Lambda)^{-1}\|\,\|V^{-1}\|
\;=\;
\kappa(V)\,\|(I-\Lambda)^{-1}\|.
\label{eq:resolvent-upper}
\end{equation}
Since $(I-\Lambda)^{-1}$ is diagonal,
$\|(I-\Lambda)^{-1}\|=\max_i|1-\lambda_i|^{-1}$, and
\begin{equation}
\|(I-\Nmat)^{-1}\|
\;\le\;
\kappa(V)\,\max_{1\le i\le d}\frac{1}{|1-\lambda_i|}.
\label{eq:resolvent-upper-bd}
\end{equation}
A matching lower bound follows from the identity
$(I-\Lambda)^{-1}=V^{-1}(I-\Nmat)^{-1}V$:
\begin{equation}
\|(I-\Nmat)^{-1}\|
\;\ge\;
\frac{1}{\kappa(V)}\,\max_{1\le i\le d}\frac{1}{|1-\lambda_i|}.
\label{eq:resolvent-lower-bd}
\end{equation}
If a dominant eigenvalue approaches unity, then
$\max_i|1-\lambda_i|^{-1}\sim(1-\rho(\Nmat))^{-1}$.
The quantity
\begin{equation}
\chi:=\frac{\kappa(V)}{1-\rho(\Nmat)}
\end{equation}
therefore provides an upper-bound-based diagnostic of the combined
effects of spectral proximity to criticality and eigenvector
non-orthogonality. It should not, in general, be interpreted as an exact
asymptotic expression for $\|(I-\Nmat)^{-1}\|$, whose realised magnitude
also depends on the alignment of the forcing and response directions
with the singular vectors of the resolvent.

\subsection{Generation operators and transient amplification}\label{sec:nonnormal:generation}

In the Galton-Watson interpretation of Sec.~\ref{sec:hawkes:branching},
$\Nmat$ governs direct offspring and $\Nmat^k$ governs expected type-to-type
transitions after $k$ generations. Under diagonalisability,
$\Nmat^k=V\Lambda^kV^{-1}$ and
\begin{equation}
\|\Nmat^k\|
\;\le\;
\|V\|\,\|\Lambda^k\|\,\|V^{-1}\|
\;=\;
\kappa(V)\,\spec(\Nmat)^k.
\label{eq:Nk-bd}
\end{equation}
For subcritical $\Nmat$, $\spec(\Nmat)^k\to 0$ as $k\to\infty$, so the
cascade decays asymptotically. However the prefactor $\kappa(V)$ can be
large; equation~(\ref{eq:Nk-bd}) accommodates substantial finite-generation
amplification before eventual exponential contraction. In branching
language, if $\bs{e}_j$ is the unit vector for a type-$j$ seed, the
expected type distribution at generation $k$ is $\Nmat^k\bs{e}_j$; certain
linear combinations of types align with the high-gain right singular
directions of $V$, leading to transiently amplified cascades before decay
sets in.

\subsection{The Jordan block case}\label{sec:nonnormal:Jordan}

If $\Nmat$ has a non-trivial Jordan block, i.e.\ an eigenvalue of
algebraic multiplicity greater than geometric multiplicity, then
$\kappa(V)$ is formally infinite (no eigenvector basis exists) and the
bound~(\ref{eq:Nk-bd}) acquires polynomial-in-$k$ corrections of the form
$k^{m-1}\spec(\Nmat)^k$, where $m$ is the largest Jordan-block size. The
qualitative phenomenon of transient amplification due to non-orthogonal
generalised eigenstructures persists, with the strict defective limit
representing the maximally non-normal case. 
Sec.~\ref{sec:physical} showed that earthquake-mechanism systems are forced by
mechanics into a \emph{near-defective} regime: $\Nmat$ is close to
upper-triangular with near-degenerate diagonal entries.

\subsection{Apparent criticality under scalar reduction}\label{sec:nonnormal:scalar}

Consider a scalar reduction of the multitype Hawkes process, for instance
by fitting a one-dimensional ETAS model to the aggregated catalogue. Such
a reduction effectively replaces $\Nmat$ by a scalar effective branching
ratio $n_\mathrm{app}$ that reproduces the observed amplification. Near
criticality, comparing the scaling
$\|(I-\Nmat)^{-1}\|\sim\kappa(V)/(1-\spec(\Nmat))$ from
(\ref{eq:resolvent-upper-bd}) with the scalar gain $1/(1-n_\mathrm{app})$
yields the heuristic identity
\begin{equation}
n_\mathrm{app}
\;\approx\;
1 - \frac{1-\spec(\Nmat)}{\kappa(V)}.
\label{eq:napp}
\end{equation}
Equation~(\ref{eq:napp}) is not strictly an identity but a
gain-matching argument; it can be made precise once a specific scalar
estimator is defined (maximum likelihood, second-moment matching, etc.).
The qualitative implication is robust: when the true $\Nmat$ is non-normal
with $\kappa(V)\gg 1$, scalar estimators applied to aggregated catalogues
\emph{systematically overestimate} the proximity to criticality. 

\subsection{Amplification of moments and covariances}\label{sec:nonnormal:moments}

Beyond the mean, non-normality also amplifies second-order statistics. Let
$d\bs{N}(t)$ denote the vector of point measures (increments of the
counting processes). Define the lagged covariance density matrix
\begin{equation}
C(\tau)\;:=\;\Cov\!\bigl(d\bs{N}(t),\,d\bs{N}(t+\tau)\bigr),
\quad \tau\in\mathbb{R},
\end{equation}
interpreted in the usual distributional sense for point processes. A
global second-order measure is the integrated covariance
$\Sigma_\mathrm{int}:=\int_{-\infty}^{+\infty}C(\tau)\,d\tau$.
For counts in a window of length $T$, the covariance of
$\bs{N}([0,T])$ scales as $T\Sigma_\mathrm{int}$ for large $T$, so
Fano factors are controlled by $\Sigma_\mathrm{int}$. Assume an
exponential memory kernel structure
\begin{equation}
\Phi(t)\;=\;B\,e^{-\beta t}\bs{1}_{t\ge 0},\quad \beta>0,
\quad B\ge 0\text{ entrywise}.
\end{equation}
The integrated branching matrix is $\Nmat=B/\beta$ and the
frequency-domain transfer function (resolvent) reads
\begin{equation}
H(\omega)\;=\;(I-\widehat{\Phi}(\omega))^{-1},\qquad
\widehat{\Phi}(\omega)\;=\;\frac{1}{1+i\omega/\beta}\,\Nmat,
\end{equation}
with $H(0)=(I-\Nmat)^{-1}$. For stationary linear Hawkes processes with
exponential kernels one obtains~\cite{BacryEtAl2013Mathematical}
\begin{equation}
\Sigma_\mathrm{int}
\;\propto\;
H(0)\,\diag(\bar{\bs{\lambda}})\,H(0)^{\top},
\label{eq:Sigma-int}
\end{equation}
showing that second-order fluctuations are amplified by the static gain
$H(0)$. Using the resolvent bound, in the spectral norm,
\begin{equation}
\|\Sigma_\mathrm{int}\|
\;\lesssim\;
\|H(0)\|^2\,\|\diag(\bar{\bs{\lambda}})\|
\;\sim\;
\frac{\kappa(V)^2}{(1-\spec(\Nmat))^2}\,\|\bar{\bs{\lambda}}\|,
\label{eq:Sigma-bd}
\end{equation}
up to multiplicative constants depending on the kernel parametrisation.
Equation~(\ref{eq:Sigma-bd}) shows that even when $\spec(\Nmat)<1$, large
non-normality inflates (i)~Fano factors (overdispersion), (ii)~cross-channel
correlations, and (iii)~sensitivity of second-order statistics to
perturbations, through the multiplicative factor $\kappa(V)^2$.

The collection of statements~(\ref{eq:resolvent-upper-bd}),
(\ref{eq:Nk-bd}), (\ref{eq:napp}), (\ref{eq:Sigma-bd}) demonstrates that
$\kappa(V)$ acts as a single scalar diagnostic of how strongly transient
amplification and observable statistics decouple from the
spectral-radius-controlled asymptotic stability. 

\subsection{Near-degenerate eigenvalues and amplification of the eigenvector condition number}
\label{sec:nonnormal:near-degenerate}

The preceding bounds show that $\kappa(V)$ measures the geometric
non-normality of the eigenbasis. We now make explicit how this quantity
can be enhanced when two eigenvalues of a three-dimensional branching
matrix become close. This mechanism is particularly relevant for the
three rupture classes considered here, since strike-slip, normal, and
reverse events may have distinct physical identities while nevertheless
possessing nearly indistinguishable triggering efficiencies.

Let
\begin{equation}
\Nmat V = V\Lambda,
\qquad
\Lambda=\diag(\lambda_1,\lambda_2,\lambda_3),
\end{equation}
with
\begin{equation}
1>\lambda_1>\lambda_2,\lambda_3,
\qquad
\lambda_2\simeq \lambda_3.
\end{equation}
All three modes are therefore subcritical, so the branching process is
asymptotically stable. Nevertheless, the conditioning of the eigenvector
matrix may become large if the two subdominant modes associated with
$\lambda_2$ and $\lambda_3$ become nearly parallel.

To isolate the effect, assume that the dominant mode remains spectrally
separated from the nearly degenerate pair:
\begin{equation}
|\lambda_1-\lambda_2|,\ |\lambda_1-\lambda_3| = O(1),
\qquad
\delta_{23}:=|\lambda_2-\lambda_3|\ll 1 .
\end{equation}
Then the possible growth of $\kappa(V)$ is controlled primarily by the
geometry of the two-dimensional eigenspace spanned by the eigenvectors
$v_2$ and $v_3$. Let the right eigenvectors be normalized as
$\|v_i\|=1$, and define the angle $\theta_{23}$ between the two
subdominant eigenvectors by
\begin{equation}
\cos\theta_{23}
:=
|v_2^{\top}v_3| .
\end{equation}
If $v_1$ remains well separated from the plane spanned by $v_2$ and
$v_3$, the leading contribution to the condition number is
\begin{equation}
\kappa(V)
\asymp
\left(
\frac{1+|\cos\theta_{23}|}
     {1-|\cos\theta_{23}|}
\right)^{1/2},
\label{eq:kappa-angle-subpair}
\end{equation}
up to a bounded prefactor depending on the orientation of $v_1$. Hence,
when the two subdominant eigenvectors almost coalesce,
$\theta_{23}\ll 1$, one obtains
\begin{equation}
\kappa(V)\asymp \frac{2}{\theta_{23}} .
\label{eq:kappa-small-angle}
\end{equation}

To justify \eqref{eq:kappa-angle-subpair}, consider first the idealized
case in which the dominant eigenvector $v_1$ is orthogonal to the plane
spanned by $v_2$ and $v_3$. Assume that the eigenvectors are normalized,
$\|v_i\|=1$. Then the Gram matrix of the eigenvector matrix
$V=(v_1,v_2,v_3)$ is
\begin{equation}
G:=V^{\top}V
=
\begin{pmatrix}
1 & 0 & 0\\
0 & 1 & v_2^{\top}v_3\\
0 & v_3^{\top}v_2 & 1
\end{pmatrix}.
\end{equation}
Writing
\begin{equation}
\rho:=|v_2^{\top}v_3|=|\cos\theta_{23}|,
\end{equation}
the two non-trivial eigenvalues of the lower $2\times2$ block of $G$ are
\begin{equation}
1+\rho,
\qquad
1-\rho.
\end{equation}
Since the singular values of $V$ are the square roots of the eigenvalues
of $G=V^{\top}V$, the largest and smallest singular values associated with
the nearly degenerate pair are
\begin{equation}
\sigma_{\max}(V)\simeq \sqrt{1+\rho},
\qquad
\sigma_{\min}(V)\simeq \sqrt{1-\rho}.
\end{equation}
Therefore,
\begin{equation}
\kappa(V)
:=
\frac{\sigma_{\max}(V)}{\sigma_{\min}(V)}
\simeq
\left(
\frac{1+\rho}{1-\rho}
\right)^{1/2}
=
\left(
\frac{1+|\cos\theta_{23}|}
     {1-|\cos\theta_{23}|}
\right)^{1/2}.
\end{equation}

More generally, if $v_1$ is not exactly orthogonal to
$\mathrm{span}\{v_2,v_3\}$ but remains uniformly separated from that
plane, the above block-diagonal Gram matrix is perturbed by bounded
off-block entries. These perturbations change the largest and smallest
singular values only by multiplicative constants independent of
$\theta_{23}$. Hence the singular behavior of $\kappa(V)$ as
$\theta_{23}\to0$ is still controlled by the nearly singular
$2\times2$ Gram block,
\begin{equation}
G_{23}
=
\begin{pmatrix}
1 & v_2^{\top}v_3\\
v_3^{\top}v_2 & 1
\end{pmatrix},
\end{equation}
and one obtains expression (\ref{eq:kappa-angle-subpair}),
where the constants implicit in $\asymp$ depend only on the orientation
of $v_1$ relative to the plane spanned by $v_2$ and $v_3$, and remain
finite as $\theta_{23}\to0$.

Thus the condition number becomes large because the eigenvectors become
nearly linearly dependent, not merely because the eigenvalues are close.
This distinction is important. For a normal matrix, eigenvectors
associated with distinct eigenvalues are orthogonal and, at exact
degeneracy, one may still choose an orthonormal basis in the degenerate
eigenspace. In that case $\kappa(V)=1$, even when
$\lambda_2=\lambda_3$. By contrast, for a non-normal matrix, two
eigenvalues can approach each other while their eigenvectors rotate
towards the same direction. The limiting case is an exceptional point:
\begin{equation}
\lambda_2=\lambda_3,
\qquad
v_2=v_3,
\end{equation}
where the matrix becomes defective and no complete eigenvector basis
exists. In this limit $\kappa(V)\to\infty$.

This behaviour can be seen explicitly by restricting $\Nmat$ to the
nearly degenerate two-mode sector. In a basis adapted to the two
subdominant modes, write the effective block as
\begin{equation}
B
=
\begin{pmatrix}
a & b\\
c & d
\end{pmatrix}.
\end{equation}
The two eigenvalues are
\begin{equation}
\lambda_{2,3}
=
\frac{a+d}{2}
\pm
\left[
\left(\frac{a-d}{2}\right)^2+bc
\right]^{1/2}.
\label{eq:lambda23-block}
\end{equation}
Their splitting is
\begin{equation}
\delta_{23}
=
|\lambda_2-\lambda_3|
=
2\left|
\left[
\left(\frac{a-d}{2}\right)^2+bc
\right]^{1/2}
\right|.
\label{eq:eigenvalue-splitting}
\end{equation}
A convenient choice of right eigenvectors is
\begin{equation}
w_{\pm}
=
\begin{pmatrix}
b\\
\lambda_{\pm}-a
\end{pmatrix},
\qquad
\lambda_{\pm}=\lambda_{2,3}.
\end{equation}
The sine of the angle between these two eigenvectors is then
\begin{equation}
\sin\theta_{23}
=
\frac{|b|\,|\lambda_2-\lambda_3|}
     {\|w_+\|\,\|w_-\|}.
\label{eq:angle-eigenvalue-splitting}
\end{equation}
Combining this relation with \eqref{eq:kappa-small-angle} gives the local
scaling
\begin{equation}
\kappa(V)
\asymp
\frac{2\|w_+\|\,\|w_-\|}
     {|b|\,|\lambda_2-\lambda_3|}
\qquad
\text{near a defective degeneracy}.
\label{eq:kappa-gap-scaling}
\end{equation}
Therefore, in the near-defective regime,
\begin{equation}
\kappa(V)\propto \frac{1}{|\lambda_2-\lambda_3|}
\label{eq:kappa-gap-scaling-boxed}
\end{equation}
up to factors that remain finite as $\lambda_2\to\lambda_3$.

Equation~\eqref{eq:kappa-gap-scaling-boxed} should not be interpreted as
a universal identity for all matrices. It describes the physically
relevant case in which near-degeneracy is accompanied by eigenvector
coalescence, as occurs close to a defective or nearly triangular
non-normal structure. If the matrix is normal, or close to normal, the
eigenvectors remain nearly orthogonal and $\kappa(V)$ remains bounded
despite the small eigenvalue gap.

For the present earthquake application, the eigenvalues are collective
triggering modes rather than individual rupture types. The three observed
types, namely strike-slip, normal, and reverse, form the physical coordinates,
whereas the eigenvectors of $\Nmat$ describe particular mixtures of these
rupture classes that reproduce themselves from one generation of
triggering to the next. The dominant eigenvalue $\lambda_1$ represents
the most persistent collective cascade mode. The two subdominant
eigenvalues $\lambda_2$ and $\lambda_3$ represent two weaker collective
triggering modes.

The regime
\begin{equation}
1>\lambda_1>\lambda_2\simeq\lambda_3
\end{equation}
therefore has the following meaning. The system is globally subcritical,
because even the largest eigenvalue satisfies $\lambda_1<1$. However, two
subdominant rupture mixtures have almost the same effective branching
strength. For example, one mode may involve a mixture dominated by
strike-slip and normal triggering, while another may involve a mixture
dominated by strike-slip and reverse triggering. If these two mixtures
have nearly identical reproductive strength and the stress-transfer
operator is non-normal, the two modal directions can become almost
parallel. The catalogue then cannot cleanly separate the two mechanisms
dynamically: small changes in stress orientation, fault geometry,
catalogue selection, or triggering asymmetry can produce large changes in
the observed mixture of rupture types.

The resulting amplification is different from ordinary proximity to
criticality. The usual critical amplification is controlled by
\begin{equation}
\frac{1}{1-\lambda_1}.
\end{equation}
Near-degeneracy of two non-normal modes produces an additional geometric
factor,
\begin{equation}
\chi
=
\frac{\kappa(V)}{1-\lambda_1}
\asymp
\frac{1}{(1-\lambda_1)|\lambda_2-\lambda_3|},
\label{eq:chi-near-degenerate}
\end{equation}
again up to bounded prefactors in the near-defective regime. Thus a
catalogue may appear much closer to criticality than suggested by the
spectral radius alone. Even if $\lambda_1$ is comfortably below one, the
small gap $|\lambda_2-\lambda_3|$ can strongly inflate $\kappa(V)$ and
therefore the gain of the multitype system.

This provides a mechanism for pseudo-criticality. The process is truly
subcritical in the branching sense:
\begin{equation}
\spec(\Nmat)=\lambda_1<1.
\end{equation}
Nevertheless, because two rupture-mechanism modes are nearly
degenerate and nearly aligned, the multitype system can exhibit large
transient amplification, enhanced covariance, and an apparent scalar
branching ratio close to one when the catalogue is aggregated. In such a
case, the apparent criticality does not arise only from
$\lambda_1\to 1^{-}$; it is amplified by the near-degenerate geometry of
the eigenvectors. The physical picture is therefore that several rupture
mechanisms are simultaneously close in their triggering properties, so
that the system has difficulty selecting a single stable modal direction.
This modal ambiguity is precisely what is measured by a large
$\kappa(V)$.

\bibliographystyle{apsrev4-2}
\bibliography{references}

\end{document}